\documentclass[doublespacing]{elsart}

\usepackage{graphicx}

\usepackage{amssymb}

\begin{document}

\hoffset = +0.10truein
\voffset = +0.10truein

\begin{frontmatter}



\title{Three-Dimensional Multi-Relaxation Time (MRT) Lattice-Boltzmann Models for Multiphase Flow}


\author{Kannan N. Premnath},
\ead{nandha@ecn.purdue.edu}
\author{John Abraham\corauthref{cor}}
\corauth[cor]{Corresponding author. }
\ead{jabraham@ecn.purdue.edu}

\address{M.J. Zucrow Labs., School of Mechanical Engineering, \\ Purdue University, West Lafayette, IN 47907.}

\begin{abstract}
In this paper, three-dimensional (3D) multi-relaxation time (MRT) lattice-Boltzmann (LB) models for multiphase
flow are presented. In contrast to the Bhatnagar-Gross-Krook (BGK) model, a widely employed kinetic model,
in MRT models the rates of relaxation processes owing to collisions of particle populations may be independently adjusted.
As a result, the MRT models offer a significant improvement in numerical
stability of the LB method for simulating fluids with lower viscosities.
We show through the Chapman-Enskog multiscale analysis that the continuum limit behavior of 3D MRT LB models
corresponds to that of the macroscopic dynamical equations for multiphase flow. We extend
the 3D MRT LB models developed to represent multiphase flow with reduced compressibility effects. The multiphase
models are evaluated by verifying the Laplace-Young relation for static drops and the frequency of oscillations of
drops. The results show satisfactory agreement with available data and significant gains in numerical stability.
\end{abstract}

\end{frontmatter}

\section{Introduction}
\label{intro}
In recent years, computational methods based on the lattice-Boltzmann equation (LBE) have
attracted much attention. They are based on the paradigm of simulating complex emergent physical
phenomena by employing minimal discrete kinetic models that represent the interactions and spatial
and temporal evolution of quasi-particles on a lattice~\cite{lbmreviews:s,lbmbooks:s}. Originally developed to
overcome certain drawbacks such as the presence of statistical noise and the lack of Galilean invariance of
lattice-gas automaton (LGA)~\cite{frisch86}, the lattice-Boltzmann equation (LBE)~\cite{mcnamara88} has
undergone a number of further refinements. They include enhanced representation of collisions of
particle populations through a relaxation process~\cite{higuera89} which was further simplified by employing
the Bhatnagar-Gross-Krook (BGK) approximation~\cite{bhatnagar54} later~\cite{qian92,chen92}. Also, its formal
connection to the Boltzmann equation, developed in the framework of non-equilibrium statistical mechanics,
was established~\cite{statconnection}.

The LBE has the beneficial feature that by incorporating physics at
scales smaller than macroscopic scales, complex fluid flows such as
multiphase flows can be simulated~\cite{luo00}. In particular, phase
segregation and interfacial fluid dynamics can be simulated by
incorporating inter-particle potentials~\cite{shanmodels}, concepts
based on free energy~\cite{swiftmodels} or the kinetic theory of
dense fluids~\cite{he98a,he99,he02}. The inter-particle potential
approach in Ref.~\cite{shanmodels}, is an earlier LBE approach for
multiphase flows and is based on non-local pseudo-potentials between
particle populations. On the other hand, free-energy based methods
are derived from thermodynamic principles that naturally provide
interfacial multiphase flow physics~\cite{swiftmodels}. The LBE
approaches based on kinetic theory of dense fluids represent physics
based on mean-field interaction forces and exclusion volume
effects~\cite{he98a,he99,he02}. More recently, LBE models have also
been developed to handle high density ratio problems, which also
overcome some of the other limitations of earlier
approaches~\cite{lee05,zheng06}. In general, the approaches above
employed the BGK approximation to represent the collision process in
the LBE and some of their applications in 3D are presented in
Ref.~\cite{3dbgkmodels}.

It is well known that the BGK model, a single-relaxation time model, often results in numerical instability when
fluids with relatively low viscosities are simulated~\cite{lallemand00}. The instability problems may be compounded in
three-dimensional (3D) flows when physics may not be adequately resolved owing to computational constraints. To address this
limitation with the standard LB models, several approaches have been proposed. In one approach, known as the entropic lattice
Boltzmann method(ELBE), the equilibrium distribution function is defined in such a way that it minimizes a certain convex function,
a Lyapunov functional designated as the $H-$function, under the constraint of local conservation laws~\cite{entropic1,entropic2}.
As a result, it ensures the positivity of the distribution function of particle populations and thus improves
the numerical stability. While this approach is endowed with elegant and desirable physical features, its numerical
accuracy is not established and it would incur relatively heavy computational overhead~\cite{yong03}.

Alternatives to the BGK model have been proposed to improve numerical stability.
In the multi-relaxation time (MRT) method~\cite{dhumieres92},
by choosing different and carefully separated time scales to represent changes in the various physical
processes due to collisions, the stability of the LBE can be significantly improved~\cite{lallemand00}.
Another interesting approach is the fractional time step method~\cite{zhang01}. In this approach,
stability is improved by considering fractional propagation of particle populations by reducing time step,
which increases computational time. While this approach was employed for the BGK model, the
underlying idea is not limited to a particular collision model. Indeed such an approach could be employed
in conjunction with the MRT model to further improve stability. Yet another possibility for improving stability
is based on the so-called two-relaxation time (TRT) models (e.g. Ref.~\cite{ginzburg05}).
These are variants of the MRT models and can be considered as limiting cases of the more general MRT models.

In addition to the computational advantage of significantly improved stability, the MRT models also have
physical advantages in that they are flexible enough to incorporate additional physics that cannot be
naturally represented by the models based on the BGK approximation. In contrast to the BGK models, MRT
models deal with the moments of the distribution functions, such as momentum and viscous stresses directly.
This moment representation provides a natural and convenient way to express various relaxation processes
due to collisions, which often occur at different time scales. Since the BGK model can represent only
a single relaxation time for all processes, it cannot naturally represent physics in certain complex
fluids. For example, as discussed in Ref.~\cite{lallemand03}, BGK models cannot represent all the
essential physics in viscoelastic flows in 3D. Moreover, MRT models make it possible to incorporate
appropriate acoustic and thermal properties through adjustable Prandtl numbers for simulating
thermo-hydrodynamics by employing a hybrid LBE/finite-difference model~\cite{lallemand03a}. BGK-based
models are not able to do so naturally. Also, in a more recent work, MRT models have been shown to be
capable of handling more general forms of diffusional transport than possible with the BGK model~\cite{rasin05}.
In addition to the capability of dealing with genuinely anisotropic diffusion problems, their MRT formulation
is able to suppress directional artifacts arising due to discrete lattice effects even in isotropic problems.
Such features are possible with MRT models because they lend themselves readily to constructing and controlling
various models to evolve at rates consistent with the dynamics of physical phenomena of interest. Indeed,
approaches based on the MRT representation are inspired by the moment method developed in the seminal work of
Maxwell~\cite{maxwell90} which was further developed by Grad~\cite{grad58}.

While prior works have demonstrated the computational advantages of the MRT models for single-phase flows in
2D by Lallemand and Luo~\cite{lallemand00} and in 3D by d'Humi\`eres \emph{et al.}~\cite{dhumieres02}
and for turbulence modeling~\cite{krafczyk03}, it has not been shown for multiphase flows in 3D in previous works.
Multiphase flows involve additional physical complexity as a result of interfacial physics involved - i.e.,
phase segregation and surface tension effects. In this case, the accuracy of the numerical discretization of
the source terms representing interfacial physics becomes an important consideration and needs to be incorporated
carefully into the framework of the MRT model. These terms should be modeled in a way that, when we establish their
relation with the various moments, the dynamical equations in the asymptotic limit should correspond to the
desired macroscopic behavior for multiphase flow. These elements were not considered in single-phase MRT models.
We introduce a second-order discretization of these source terms and avoid implicitness through a transformation
which, to our best knowledge, is applied to MRT models for multiphase flows in 3D for the first time in this work. Recently,
a MRT model for 2D multiphase problems was developed~\cite{mccracken05}. There are significant differences
in the development and implementation of 2D and 3D MRT models. These will be discussed below. Furthermore,
realistic multiphase flows are 3D in nature and in this work, we develop a 3D MRT model for such problems by
employing a model for interfacial physics based on the kinetic theory of dense fluids~\cite{he98a,he99,he02}.

The specific contributions of this work will now be stated. We develop a 3D MRT model incorporating phase segregation
and surface tension effects through source terms discretized by using trapezoidal rule integration and
simplified through a transformation to achieve an effectively explicit 3D MRT model for multiphase flows.
Since the underlying lattice structure for 2D and 3D models are different, the moment basis for the
corresponding MRT models are different, and is more complicated in 3D. We present the theoretical developments
based on the 3D moment basis for multiphase flow which, to our best knowledge, has not been considered in prior
work. We derive the continuum equations for multiphase flow from the 3D MRT model through a Chapman-Enskog
analysis~\cite{chapman64}, provide the dynamical relations between the various moments and forcing terms
representing interfacial physics including surface tension forces, develop the relationships between
the gradients of momentum fields in terms of the non-equilibrium parts of certain moments and their
corresponding relaxation times, and explicitly derive the relationship between the transport coefficients
for the fluid flow and appropriate relaxation times. To improve the stability of the approach, the model
is then transformed in such a way that compressibility effects are reduced. Furthermore, we evaluate the
accuracy and gains in stability of the MRT model for some canonical multiphase problems in 3D.

As the MRT models are endowed with greater computational stability and potential to incorporate additional
physics, there is considerable interest in their applications to multiphase problems: The 2D MRT model
developed by McCracken and Abraham~\cite{mccracken05} extended for axisymmetric problems using the
axisymmetric LB model~\cite{premnath05} has been employed to study the physics of break up of liquid jets~\cite{mccracken05a}.
This axisymmetric MRT model and the 3D MRT model developed in this work have also been employed to study the
physics of head-on as well as off-center binary drop collisions, respectively~\cite{premnath05a}. The physically inspired
MRT approach developed in this paper also provides a natural framework for incorporating additional physics such
as viscoelastic or thermal effects in multiphase flows through an LB model, which are subjects of future work.
The rest of the paper is organized as follows. In Section~\ref{mrt3model}, the 3D MRT LBE multiphase models are developed.
In Section~\ref{mrt3macrod}, the macroscopic dynamical equations of these models are derived by using a Chapman-Enskog multiscale
analysis. Section~\ref{mrt3modelc} transforms the model to simulate multiphase flow with reduced compressibility effects.
In Section~\ref{mrt3benchmarks}, the model is applied to benchmark problems to evaluate its accuracy and gains in stability. Finally, the paper
closes with summary and conclusions in Section~\ref{summary}.

\section{3D MRT LBE Model for Multiphase Flow}
\label{mrt3model}
We develop a MRT model for multiphase flows in which the underlying interfacial physics is based on the kinetic
theory of dense fluids. In particular, we consider that the particle populations representing the dense fluids
experience mean-field interaction forces and respect Enskog effects for dense non-ideal fluids~\cite{he98a,he99}.
The effect of collisions of particle populations is represented though a generalized relaxation process in which
the distribution functions for discrete velocity directions approach their corresponding local equilibrium values
at characteristic time scales given in terms of a generalized collision or scattering matrix.

\begin{figure}
\begin{center}
\includegraphics[height=3.25in,clip=]{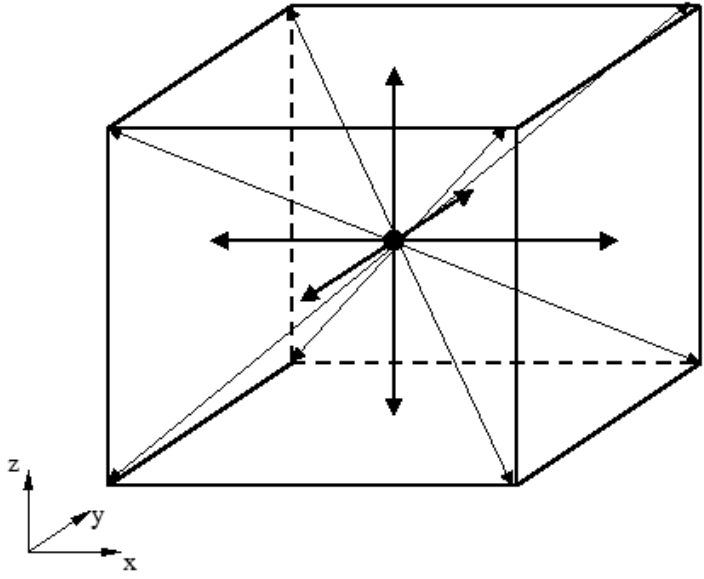}
\caption{D3Q15 lattice.}
\label{fig:d3q15mrt}
\end{center}
\end{figure}

\begin{figure}
\begin{center}
\includegraphics[height=3.25in,clip=]{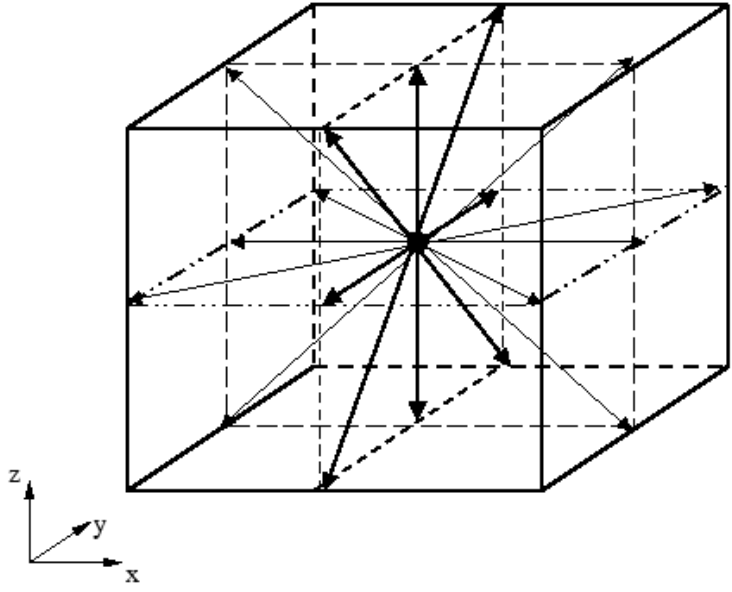}
\caption{D3Q19 lattice.}
\label{fig:d3q19mrt}
\end{center}
\end{figure}

In the following, we consider subscripts with Greek symbols for particle velocity directions and Latin symbols for Cartesian components of
spatial directions. We assume summation convention for repeated indices for the components of spatial directions.
In addition, unless otherwise stated, we follow the convention that vectors corresponding to three-dimensional position
space are represented by non-capitalized symbols with arrowheads; the vectors with a particle velocity basis are denoted
by non-capitalized boldface symbols; the square matrices constructed from the particle velocity basis are represented by
non-boldface capitalized symbols. We consider the following MRT LBE with a source term that gives rise to phase segregation and
surface tension effects:
\begin{eqnarray}
f_{\alpha}( \overrightarrow{x}+ \overrightarrow{e_{\alpha}}\delta_t,t+\delta_t )-f_{\alpha}( \overrightarrow{x},t )&=&
\Omega_{\alpha}|_{(x,t)}+\nonumber\\
& &
\frac{1}{2}\left[S_{\alpha}|_{(x,t)}+
S_{\alpha}|_{(x+e_{\alpha}\delta_t,t+\delta_t)}
\right]\delta_t.
\label{eq:mrt3lbe}
\end{eqnarray}
Here, $f_{\alpha}$ is the discrete single-particle distribution function, corresponding to the particle velocity, $\overrightarrow{e_{\alpha}}$,
where $\alpha$ is the velocity direction. The Cartesian component of the particle velocity $c$,
is given by $c=\delta_x/\delta_t$, where $\delta_x$ is the lattice spacing and $\delta_t$ is the time step.
The left hand side (LHS) of this equation represents the change in the distribution function as particle populations advect from one lattice node to its
adjacent one along the characteristic direction represented by the discrete lattice velocity direction $\overrightarrow{e_{\alpha}}$
(see Figs.~\ref{fig:d3q15mrt} and ~\ref{fig:d3q19mrt}). On the other hand, the right hand side (RHS) represents the effect of particle
collisions and force interactions.

The form of the collision term $\Omega_{\alpha}$ that incorporates MRT collision processes will be discussed below. The source term $S_{\alpha}$
which models the interfacial physics in multiphase flow needs to be accurately represented. In Eq. (\ref{eq:mrt3lbe}), we considered a second-order
trapezoidal rule discretization of this term. It may be written as~\cite{he98a}
\begin{equation}
S_{\alpha}=\frac{(e_{\alpha j}-u_j)(F_j^{I}+F_{ext,j})}{\rho RT} f_{\alpha}^{eq,M} (\rho,\overrightarrow{u}),
\label{eq:source3mrt}
\end{equation}
where $F_j^{I}$ is the interaction force term that models the interfacial physics, while
$F_{ext,j}$ corresponds to external or imposed forces such as gravity. $F_j^{I}$ is modeled as a function of density following
the work of van der Waals~\cite{rowlinson82}. The exclusion-volume effect of Enskog~\cite{chapman64} is also incorporated to
account for increase in collision probability due to the increase in the density of non-ideal fluids. These features account
for the phase segregation and surface tension effects. For more details, the reader is referred to Ref.~\cite{he98a}.
In Eq. (\ref{eq:source3mrt}), $f_{\alpha}^{eq,M}$ is the local discrete Maxwellian and its functional expression will be presented below.
Effectively, $F_j^{I}$ may be written as
\begin{equation}
F_j^{I}=-\partial_j \psi + F_{s,j}.
\label{eq:force3mrt}
\end{equation}
In Eq. (\ref{eq:force3mrt}), $F_{s,j}$ represents the surface tension force and is related
to the density $\rho$ and its gradients by
\begin{equation}
F_{s,j}=\kappa \rho \partial_j \nabla^2 \rho,
\label{eq:forcempmrt3}
\end{equation}
where $\kappa$ is a surface tension parameter. It is related to the surface tension $\sigma$ of the
fluid by the equation~\cite{evans79}
\begin{equation}
\sigma=\kappa \int \left( \frac{\partial \rho}{\partial n} \right)^2 dn,
\label{eq:surft}
\end{equation}
where $n$ is the direction normal to the interface. Thus, the surface tension is a function of
both the parameter $\kappa$ and the density profile across the interface.

The term $\psi$ in Eq. (\ref{eq:force3mrt}) refers to the non-ideal part of the equation of state (EOS)
\begin{equation}
\psi(\rho)=P-\rho R T.
\label{eq:nonideal}
\end{equation}
In this work, the Carnahan-Starling-van der Waals EOS~\cite{carnahan69},
\begin{equation}
P=\rho R T \left\{ \frac{1+\gamma+\gamma^2-\gamma^3}{(1-\gamma)^3}\right\} -a\rho^2,
\end{equation}
is employed where $\gamma=b\rho/4$. The parameter $a$ is related to the inter-particle pair-wise
potential and $b$ is related to the effective diameter $d$ of the particle, and the mass $m$ of a single
particle, by $b=2\pi d^3/3m$. $\psi$ assumes an important role in determining phase segregation~\cite{he98a}.
The Carnahan-Starling-van der Waals EOS has a  $P-1/\rho-T$ curve, in which $dP/d\rho<0$ for certain range
of values of $\rho$, when the fluid temperature is below its critical value. This part of the
curve represents an unstable physical situation and is the driving mechanism responsible for keeping
the phases segregated and in maintaining a \emph{self-generated} interface that is diffuse with a thickness of
about $3$-$4$ lattice grid points.

The local discrete Maxwellian $f_{\alpha}^{eq,M}(\rho,\overrightarrow{u})$ in Eq. (\ref{eq:source3mrt}) is
obtained from a truncated expansion in terms of fluid velocity $\overrightarrow{u}$ of its continuous version, after approximating
it for discrete particle velocities $\overrightarrow{e_{\alpha}}$~\cite{statconnection}. As a result, it is a function of the fluid
densities and velocities and is given by
\begin{equation}
f_{\alpha}^{eq,M}(\rho,\overrightarrow{u})=
\omega_{\alpha} \rho \left\{
1+\frac{\overrightarrow{e_{\alpha}} \cdotp \overrightarrow{u} }{RT}+
  \frac{\left( \overrightarrow{e_{\alpha}} \cdotp  \overrightarrow{u} \right)^2}{2(RT)^2}-
  \frac{1}{2}\frac{ \overrightarrow{u} \cdotp \overrightarrow{u} }{RT}
\right\},
\label{eq:trunceqmrt3}
\end{equation}
where $w_{\alpha}$ is the weighting factor. In this representation of the local Maxwellian, the factor $RT$
is related to the speed of sound $c_{s}$ of the model through $RT=c_{s}^2$, where $c_{s}=1/\sqrt{3}c$.
For the three-dimensional, fifteen-velocity
(D3Q15) model~\cite{qian92}, shown in Fig.~\ref{fig:d3q15mrt}, the weighting factors become
\begin{equation}
\omega_{\alpha}=\left\{\begin{array}{ll}
   {\frac{2}{9}}&{ \alpha=1}\\
   {\frac{1}{9}}&{ \alpha=2,\cdots,7}\\
   {\frac{1}{72}}&{ \alpha=9,\cdots,15,}
\end{array} \right.
\label{eq:weightd3q15}
\end{equation}
and for the three-dimensional, nineteen-velocity (D3Q19) model~\cite{qian92}, shown in Fig.~\ref{fig:d3q19mrt}, we have
\begin{equation}
\omega_{\alpha}=\left\{\begin{array}{ll}
   {\frac{1}{3}}&{ \alpha=1}\\
   {\frac{1}{18}}&{ \alpha=2,\cdots,7}\\
   {\frac{1}{36}}&{ \alpha=8,\cdots,19.}
\end{array} \right.
\label{eq:weightd3q19}
\end{equation}
The corresponding particle velocity directions are
\begin{equation}
\overrightarrow{e_{\alpha}} = \left\{\begin{array}{ll}
   {(0,0,0)}&{ \alpha=1}\\
   {(\pm 1,0,0),(0,\pm 1,0),(0,0,\pm 1)}&{ \alpha=2,\cdots,7}\\
   {(\pm 1,\pm 1,\pm 1)}&{ \alpha=8,\cdots,15,}
\end{array} \right.
\label{eq:velocityd3q15}
\end{equation}
and
\begin{equation}
\overrightarrow{e_{\alpha}} = \left\{\begin{array}{ll}
   {(0,0,0)}&{ \alpha=1}\\
   {(\pm 1,0,0),(0,\pm 1,0),(0,0,\pm 1)}&{ \alpha=2,\cdots,7}\\
   {(\pm 1,\pm 1,0),(\pm 1,0,\pm 1),(0,\pm 1,\pm 1)}&{ \alpha=8,\cdots,19,}
\end{array} \right.
\label{eq:velocityd3q19}
\end{equation}
respectively.

We express the effect of collisions in Eq. (\ref{eq:mrt3lbe}) as a relaxation process through a multi-relaxation
time (MRT) model for the collision term as shown below~\cite{lallemand00,dhumieres92,dhumieres02}:
\begin{equation}
\Omega_{\alpha}=-\sum_{\beta}\Lambda_{\alpha \beta}\left( f_{\beta}-f_{\beta}^{eq} \right),
\label{eq:mrt3coll}
\end{equation}
where $\Lambda_{\alpha \beta}$ is the component of the collision or the scattering matrix $\Lambda$.
Here, the equilibrium distribution $f_{\beta}^{eq}$ is an appropriate function of the conserved moments
such as the density and momentum and is, in general, not necessarily the local discrete Maxwellian given
by Eq. (\ref{eq:trunceqmrt3}). The hydrodynamic field variables are obtained by taking the kinetic moments
of the distribution functions as
\begin{eqnarray}
\rho&=&\sum_{\alpha} f_{\alpha},\\
\rho u_i&=&\sum_{\alpha} f_{\alpha} e_{\alpha i}.
\end{eqnarray}

Equation (\ref{eq:mrt3lbe}) is implicit. For computational convenience, it can be made explicit if
we introduce the transformation~\cite{he98a}
\begin{equation}
\bar{f}_{\alpha}=f_{\alpha}-
\frac{1}{2}S_{\alpha}\delta_t.
\label{eq:implicittr1mrt3}
\end{equation}
As a result, Eq. (\ref{eq:mrt3lbe}) is replaced by the following effectively explicit scheme:
\begin{eqnarray}
\bar{f}_{\alpha}(\overrightarrow{x}+\overrightarrow{e_{\alpha}}\delta_t,t+\delta_t )-
\bar{f}_{\alpha}(\overrightarrow{x} ,t )&=&
- \sum_{\beta} \Lambda_{\alpha \beta}\left( \bar{f}_{\beta}-f_{\beta}^{eq} \right)|_{(x,t)}+ \nonumber\\
& &
\sum_{\beta} \left(I_{\alpha\beta}-\frac{1}{2}\Lambda_{\alpha\beta}\right)S_{\beta}|_{(x,t)}\delta_t,
\label{eq:mrt3lbe2}
\end{eqnarray}
where $I_{\alpha\beta}$ is the component of the identity matrix $ \mathcal{I} $.

Next, we construct appropriate sets of linearly independent moments from
the distribution functions in velocity space. Since moments of the distribution
function, such as the momentum and viscous stresses, directly represent physical
quantities, the \emph{moment representation} offers a natural and convenient way to
express the relaxation processes due to collisions. One particular advantage of this
representation is that the time scales of the various processes represented in terms of
moments can be controlled independently. The moments are constructed from the distribution
function through a transformation matrix $ \mathcal{T} $ comprising a linearly independent
set of vectors, i.e.
\begin{equation}
\widehat{\bar{\mbox{\boldmath$f$}}}= \mathcal{T} \bar{\mbox{\boldmath$f$}},
\label{eq:transvmmrt3}
\end{equation}
where
\begin{equation}
\bar{\mbox{\boldmath$f$}}=\left[\bar{f}_1,\bar{f}_2,\bar{f}_3,\bar{f}_4,\bar{f}_5,\bar{f}_6,
                \bar{f}_7,\bar{f}_8,\bar{f}_9,\bar{f}_{10},\bar{f}_{11},
                \bar{f}_{12},\bar{f}_{13},\bar{f}_{14},\bar{f}_{15}
              \right]^{T}
\label{eq:distd3q15}
\end{equation}
is the vector representing the particle distribution functions for the D3Q15 model and
\begin{eqnarray}
\bar{\mbox{\boldmath$f$}}&=&\left[\bar{f}_1,\bar{f}_2,\bar{f}_3,\bar{f}_4,\bar{f}_5,\bar{f}_6,
                \bar{f}_7,\bar{f}_8,\bar{f}_9,\bar{f}_{10},\bar{f}_{11},
                \bar{f}_{12},\bar{f}_{13},\right. \nonumber\\
                     & & \left. \bar{f}_{14},\bar{f}_{15},\bar{f}_{16},\bar{f}_{17},\bar{f}_{18},\bar{f}_{19}
              \right]^{T}
\label{eq:distd3q19}
\end{eqnarray}
for the D3Q19 model.

In Eqs. (\ref{eq:transvmmrt3}) - (\ref{eq:distd3q19}), the superscript $'T'$ represents transpose
of a matrix and the vector $\widehat{\bar{\mbox{\boldmath$f$}}}$ represents the moments
\begin{equation}
\widehat{\bar{\mbox{\boldmath$f$}}}=
\left[\rho,e,e^2,j_x,q_x,j_y,q_y,j_z,q_z,3p_{xx},p_{ww},
      p_{xy},p_{yz},p_{zx},m_{xyz}\right]^{T},
\label{eq:momentd3q15}
\end{equation}
and
\begin{eqnarray}
\widehat{\bar{\mbox{\boldmath$f$}}}&=&
\left[\rho,e,e^2,j_x,q_x,j_y,q_y,j_z,q_z,3p_{xx},3\pi_{xx},p_{ww},\right.\nonumber\\
      & & \left.
      \pi_{ww},p_{xy},p_{yz},p_{zx},m_x,m_y,m_z\right]^{T},
\label{eq:momentd3q19}
\end{eqnarray}
for the D3Q15 and D3Q19 models, respectively.
For the D3Q15 model, $e$ and $e^2$ represent kinetic energy that is independent of density and square of energy, respectively,
$j_x$, $j_y$ and $j_z$ are the components of the momentum or mass flux $(j_x=\rho u_x, j_y=\rho u_y, j_z=\rho u_z)$,
$q_x$, $q_y$, $q_z$ are the components of the energy flux, and $p_{xx}$,
$p_{xy}$, $p_{yz}$ and $p_{zx}$ are the components of the symmetric traceless viscous
stress tensor. The other two normal components of the viscous stress tensor, $p_{yy}$ and $p_{zz}$,
can be constructed from $p_{xx}$ and $p_{ww}$, where
$p_{ww}=p_{yy}-p_{zz}$. The quantity $m_{xyz}$ is an antisymmetric third-order moment. For the D3Q19 model, instead of
the moment $m_{xyz}$, we have five additional moments: $3\pi_{xx},3\pi_{ww},m_x,m_y$ and $m_z$. The first two of these
moments have the same symmetry as the diagonal part of the traceless viscous tensor $p_{ij}$, while the last three vectors
are parts of a third rank tensor, with the symmetry of $j_k p_{mn}$~\cite{dhumieres02}.

The underlying principle for the construction of the transformation matrix is based on the observation
that the collision matrix reduces to a diagonal form in an appropriate orthonormal basis, which is obtained
by combinations of monomials of Cartesian components of the particle velocity directions $\overrightarrow{e_{\alpha}}$~\cite{dhumieres92}.
Following d'Humieres \emph{et al.}~\cite{dhumieres02}, for the D3Q15 model we introduce a
transformation matrix $ \mathcal{T} $, which represents components of the $15$ orthogonal
basis column vectors $\mbox{\boldmath$v$}_{\beta}$ that form the moment basis, i.e.
\begin{equation}
\mathcal{T}=
\left[\mbox{\boldmath$v$}_1, \mbox{\boldmath$v$}_2,\mbox{\boldmath$v$}_3,
      \mbox{\boldmath$v$}_4, \mbox{\boldmath$v$}_5,\mbox{\boldmath$v$}_6,
      \mbox{\boldmath$v$}_7, \mbox{\boldmath$v$}_8,\mbox{\boldmath$v$}_9,
      \mbox{\boldmath$v$}_{10}, \mbox{\boldmath$v$}_{11},\mbox{\boldmath$v$}_{12},
      \mbox{\boldmath$v$}_{13}, \mbox{\boldmath$v$}_{14},\mbox{\boldmath$v$}_{15}
\right]^{T},
\label{eq:trmatrixd3q15}
\end{equation}
where the components of each column vector may be written as
$v_{1 \alpha}   =   | \overrightarrow{e_{\alpha}} |^{0};$
$v_{2 \alpha}   =   | \overrightarrow{e_{\alpha}} |^{2}-2;$
$v_{3 \alpha}   =   \frac{1}{2}\left( 15| \overrightarrow{e_{\alpha}} |^{4}-
           55| \overrightarrow{e_{\alpha}} |^{2}+32 \right);$
$v_{4 \alpha}   =   e_{\alpha x};$
$v_{5 \alpha}   =   \frac{1}{2}\left( 5| \overrightarrow{e_{\alpha}} |^{2}-13 \right)e_{\alpha x};$
$v_{6 \alpha}   =   e_{\alpha y};$
$v_{7 \alpha}   =   \frac{1}{2}\left( 5| \overrightarrow{e_{\alpha}} |^{2}-13 \right)e_{\alpha y};$
$v_{8 \alpha}   =   e_{\alpha z};$
$v_{9 \alpha}   =   \frac{1}{2}\left( 5| \overrightarrow{e_{\alpha}} |^{2}-13 \right)e_{\alpha z};$
$v_{10 \alpha}   =    3e_{\alpha x}^2-| \overrightarrow{e_{\alpha}} |^{2};$
$v_{11 \alpha}   =    e_{\alpha y}^2-e_{\alpha z}^2;$
$v_{12 \alpha}   =    e_{\alpha x}e_{\alpha y};$
$v_{13 \alpha}   =    e_{\alpha y}e_{\alpha z};$
$v_{14 \alpha}   =    e_{\alpha x}e_{\alpha z};$
$v_{15 \alpha}   =    e_{\alpha x}e_{\alpha x}e_{\alpha z}.$

For the D3Q19 model, we have
\begin{eqnarray}
\mathcal{T}&=&
\left[\mbox{\boldmath$v$}_1, \mbox{\boldmath$v$}_2,\mbox{\boldmath$v$}_3,
      \mbox{\boldmath$v$}_4, \mbox{\boldmath$v$}_5,\mbox{\boldmath$v$}_6,
      \mbox{\boldmath$v$}_7, \mbox{\boldmath$v$}_8,\mbox{\boldmath$v$}_9,
      \mbox{\boldmath$v$}_{10}, \mbox{\boldmath$v$}_{11},\mbox{\boldmath$v$}_{12},\mbox{\boldmath$v$}_{13}, \right.\nonumber\\
      & & \left.
      \mbox{\boldmath$v$}_{14},\mbox{\boldmath$v$}_{15},
      \mbox{\boldmath$v$}_{16}, \mbox{\boldmath$v$}_{17},\mbox{\boldmath$v$}_{18},
      \mbox{\boldmath$v$}_{19}
\right]^{T}
\label{eq:trmatrixd3q19}
\end{eqnarray}
based on $19$ orthogonal basis column vectors. The components of each of these vectors are
$v_{1 \alpha}   =   |  \overrightarrow{e_{\alpha}} |^{0};$
$v_{2 \alpha}   =   19|  \overrightarrow{e_{\alpha}} |^{2}-30;$
$v_{3 \alpha}   =   \frac{1}{2}\left( 21|  \overrightarrow{e_{\alpha}} |^{4}-
                   53|  \overrightarrow{e_{\alpha}} |^{2}+24 \right);$
$v_{4 \alpha}   =   e_{\alpha x};$
$v_{5 \alpha}   =   \left( 5|  \overrightarrow{e_{\alpha}} |^{2}-9 \right)e_{\alpha x};$
$v_{6 \alpha}   =   e_{\alpha y};$
$v_{7 \alpha}   =   \left( 5|  \overrightarrow{e_{\alpha}} |^{2}-9 \right)e_{\alpha y};$
$v_{8 \alpha}   =   e_{\alpha z};$
$v_{9 \alpha}   =   \left( 5|  \overrightarrow{e_{\alpha}} |^{2}-9 \right)e_{\alpha z};$
$v_{10 \alpha}   =    3e_{\alpha x}^2-|e_{\alpha}|^{2};$
$v_{11 \alpha}   =    \left(3|  \overrightarrow{e_{\alpha}} |^{2}-5\right)
             \left(3e_{\alpha x}^2-|  \overrightarrow{e_{\alpha}} |^{2}\right);$
$v_{12 \alpha}   =    e_{\alpha y}^2-e_{\alpha z}^2;$
$v_{13 \alpha}   =    \left(3|  \overrightarrow{e_{\alpha}} |^{2}-5\right)
             \left(e_{\alpha y}^2-e_{\alpha z}^2\right);$
$v_{14 \alpha}   =    e_{\alpha x}e_{\alpha y};$
$v_{15 \alpha}   =    e_{\alpha y}e_{\alpha z};$
$v_{16 \alpha}   =    e_{\alpha x}e_{\alpha z};$
$v_{17 \alpha}   =    \left(e_{\alpha y}^2-e_{\alpha z}^2\right)e_{\alpha x};$
$v_{18 \alpha}   =    \left(e_{\alpha z}^2-e_{\alpha x}^2\right)e_{\alpha y};$
$v_{19 \alpha}   =    \left(e_{\alpha x}^2-e_{\alpha y}^2\right)e_{\alpha z}.$

These matrices are formed by a set of linearly independent orthogonal basis vectors
(i.e. $\mbox{\boldmath$v$}_{\alpha}\cdot\mbox{\boldmath$v$}_{\beta}=l_{\alpha}\delta_{\alpha\beta}$,
where $l_{\alpha}$ is a normalizing constant) which are constructed by a Gram-Schmidt procedure such that they
diagonalize the collision matrix $\Lambda$ i.e.
$\widehat{\Lambda}= \mathcal{T} \Lambda \mathcal{T}^{-1}$
~\cite{dhumieres02}, where the matrix $\widehat{\Lambda}$ in moment space is a diagonal matrix.
This orthogonal set is built in increasing order of moments and then arranged in increasing order
of complexity of the tensorial representation of the moments. An example of such a construction with some
details for a 2D MRT model is given by Bouzidi \emph{et al.}~\cite{bouzidi01}.
As an example, for unit lattice spacing and time steps, i.e. $c=1$, the simplified form of the transformation matrices
is given in the Appendix.

The equilibrium distribution functions $\widehat{{\mbox{\boldmath$f$}}}^{eq}$ in moment space are related to those
in velocity space, $\mbox{\boldmath$f$}^{eq}$, through the transformation matrix
\begin{eqnarray}
\widehat{{\mbox{\boldmath$f$}}}^{eq}&=&   \mathcal{T} \mbox{\boldmath$f$}^{eq} \nonumber\\
&=&
\left[\rho,e^{eq},e^{2,eq},j_x,q_x^{eq},j_y,q_y^{eq},j_z,q_z^{eq},3p_{xx}^{eq},p_{ww}^{eq},
      p_{xy}^{eq},p_{yz}^{eq},p_{zx}^{eq},m_{xyz}^{eq}\right]^{T}
\label{eq:momenteqd3q15}
\end{eqnarray}
for the D3Q15 model and
\begin{eqnarray}
\widehat{{\mbox{\boldmath$f$}}}^{eq}&=&  \mathcal{T} \mbox{\boldmath$f$}^{eq} \nonumber\\
&=&
\left[\rho,e^{eq},e^{2,eq},j_x,q_x^{eq},j_y,q_y^{eq},j_z,q_z^{eq},3p_{xx}^{eq},3\pi_{ww}^{eq},p_{ww}^{eq},
      \pi_{ww}^{eq},\right.\nonumber\\
& &\left. p_{xy}^{eq},p_{yz}^{eq},p_{zx}^{eq},m_x^{eq},m_y^{eq},m_z^{eq}\right]^{T}
\label{eq:momenteqd3q19}
\end{eqnarray}
for the D3Q19 model. Notice that for the conserved or hydrodynamic moments corresponding to density and components of  momentum, i.e.
$\widehat{\bar{f}}_{\beta}$, where $\beta=1,4,6,8$, their equilibrium distributions in moment space are also the
same, i.e. $\widehat{f}^{eq}_{\beta}=\widehat{\bar{f}}_{\beta}$, where $\beta=1,4,6,8$
since the collision process does not alter hydrodynamic moments.

The expression for the equilibrium distribution functions
of the non-conserved or kinetic moments, which are in turn algebraic functions of the conserved moments and
obtained by optimizing isotropy and Galilean invariance, are given by
~\cite{dhumieres02}
$e^{eq}=-\rho+\frac{\left( j_x^2+j_y^2+j_z^2 \right)}{\rho};$
$e^{2,eq}=\rho-5\frac{\left( j_x^2+j_y^2+j_z^2 \right)}{\rho};$
$q_x^{eq}=-\frac{7}{3} j_x;$
$q_y^{eq}=-\frac{7}{3} j_y;$
$q_z^{eq}=-\frac{7}{3} j_z;$
$p_{xx}^{eq}=\frac{1}{3}\frac{\left[2j_x^2-(j_y^2+j_z^2) \right]}{\rho};$
$p_{ww}^{eq}=\frac{\left[ j_y^2-j_z^2 \right]}{\rho};$
$p_{xy}^{eq}=\frac{ j_x j_y }{\rho};$
$p_{yz}^{eq}=\frac{ j_y j_z }{\rho};$
$p_{xz}^{eq}=\frac{ j_x j_z }{\rho};$
$m_{xyz}^{eq}=0$

for the D3Q15 model and
$e^{eq}=-11\rho+19\frac{\left( j_x^2+j_y^2+j_z^2 \right)}{\rho};$
$e^{2,eq}=3\rho-\frac{11}{2}\frac{\left( j_x^2+j_y^2+j_z^2 \right)}{\rho};$
$q_x^{eq}=-\frac{2}{3} j_x;$
$q_y^{eq}=-\frac{2}{3} j_y;$
$q_z^{eq}=-\frac{2}{3} j_z;$
$p_{xx}^{eq}=\frac{1}{3}\frac{\left[2j_x^2-(j_y^2+j_z^2) \right]}{\rho};$
$\pi_{xx}^{eq}=-\frac{1}{2}p_{xx}^{eq};$
$p_{ww}^{eq}=\frac{\left[ j_y^2-j_z^2 \right]}{\rho};$
$\pi_{ww}^{eq}=-\frac{1}{2}p_{ww}^{eq};$
$p_{xy}^{eq}=\frac{ j_x j_y }{\rho};$
$p_{yz}^{eq}=\frac{ j_y j_z }{\rho};$
$p_{xz}^{eq}=\frac{ j_x j_z }{\rho};$
$m_x^{eq}=0;$
$m_y^{eq}=0;$
$m_z^{eq}=0$

for the D3Q19 model.

The collision matrix $\widehat{\Lambda}$ in the moment space is given by
\begin{equation}
\widehat{\Lambda}=
diag\left[s_1,s_2,s_3,s_4,s_5,s_6,s_7,s_8,s_9,s_{10},s_{11},s_{12},s_{13},s_{14},s_{15}\right]^{T}
\label{eq:collmatd3q15}
\end{equation}
for the D3Q15 model and
\begin{eqnarray}
\widehat{\Lambda}&=&
diag\left[s_1,s_2,s_3,s_4,s_5,s_6,s_7,s_8,s_9,s_{10},s_{11},s_{12},s_{13},\right.\nonumber\\
     & & \left.
          s_{14},s_{15},s_{16},s_{17},s_{18},s_{19}\right]^{T}
\label{eq:collmatd3q19}
\end{eqnarray}
for the D3Q19 model,
where the parameters $s_{\beta}$ represent the inverse of the relaxation times of the various
moments $\widehat{{\mbox{\boldmath$\bar{f}$}}}$ in reaching their equilibrium values $\widehat{{\mbox{\boldmath$f$}}}^{eq}$.

The variables $s_1$, $s_4$, $s_6$ and $s_8$ are the relaxation
parameters corresponding to the collision invariants $\rho$, $j_x$,
$j_y$ and $j_z$, respectively. Since the collision process conserves
these particular moments, the choice of their corresponding
relaxation times is immaterial. This is because they are each
specified to relax during collisions to their local equilibrium
which are actually defined to be the corresponding pre-collision
value of the respective quantities in the equilibrium distribution
$f^{eq}$ (see Eqs. (\ref{eq:momenteqd3q15}) and
(\ref{eq:momenteqd3q19})). So, whatever be the relaxation parameters
for these quantities, the collision process does not change their
values. The relaxation parameters for these collision invariants
could be set to zero when there is no forcing term $S_{\alpha}$ in
the LBE. However, care needs to be exercised in selecting the
relaxation parameters of MRT LBE with forcing term. This has been
pointed out in Ref.~\cite{ginzbourg94}. This is because the
collision matrix also influences the forcing term in the effectively
explicit MRT LBE (see Eq. (\ref{eq:mrt3lbe2})). To correctly obtain
the continuum asymptotic limit of this LBE, i.e. the macroscopic
hydrodynamical equations for multiphase flow, derived in the next
section, the relaxation times for the conserved moments should be
set to non-zero values. For simplicity, we have chosen a value of
unity as the relaxation parameter for these moments
($s_1=s_4=s_6=s_8=1$) that preserves the influence of the
second-order discretization of the source terms in Eq.
(\ref{eq:mrt3lbe}) (see also Ref.~\cite{mccracken05}). Furthermore,
for the D3Q15 model, it will be shown through the Chapman-Enskog
analysis in Section~\ref{mrt3macrod} that the parameter $s_2$ is
related to the bulk viscosity and $s_{10}$ through $s_{14}$ are all
related to the kinematic viscosity. This leaves us with $s_3$,
$s_5$, $s_7$, $s_9$, and $s_{15}$ as free parameters in this model.
On the other hand, for the D3Q19 model by analogy, the parameter
$s_2$ is related to the bulk viscosity and $s_{10}$, $s_{12}$,
$s_{14}$ through $s_{16}$ are each related to the kinematic
viscosity. Then, $s_3$, $s_5$, $s_7$, $s_9$, $s_{11}$, $s_{13}$,
$s_{17}$ through $s_{19}$ are the free parameters.

To obtain the LBE model in moment space, the source terms are also transformed as follows:
\begin{equation}
\widehat{\mbox{\boldmath$\varsigma$}}= \mathcal{T} \mbox{\boldmath$\varsigma$},
\end{equation}
where $\mbox{\boldmath$\varsigma$}$ is the column vector corresponding to the components of the source term $S_{\alpha}$;
$\widehat{\mbox{\boldmath$\varsigma$}}$ corresponds to the components of $\widehat{S}_{\alpha}$.
Now, pre-multiplying Eq. (\ref{eq:mrt3lbe2}) by the transformation matrix $\mathcal{T}$, we obtain the
3D MRT model
\begin{eqnarray}
& &\widehat{\bar{f}}_{\alpha}( \overrightarrow{x} + \overrightarrow{e_{\alpha}} \delta_t,t+\delta_t )-
\widehat{\bar{f}}_{\alpha}( \overrightarrow{x},t )= \nonumber \\
& &
-\sum_{\beta}  \widehat{\Lambda}_{\alpha \beta}\left( \widehat{\bar{f}}_{\beta}-\widehat{f}_{\beta}^{eq} \right)|_{(x,t)}+
 \sum_{\beta}  \left(I_{\alpha\beta}-\frac{1}{2}\widehat{\Lambda}_{\alpha\beta}\right) \widehat{S}_{\beta}|_{(x,t)}\delta_t.
 \label{eq:mrt3lbe3}
 \end{eqnarray}

The hydrodynamic field variables can be obtained from the distribution functions in velocity space as
\begin{eqnarray}
\rho&=&\sum_{\alpha} \bar{f}_{\alpha},\\
\rho u_i&=&\sum_{\alpha} \bar{f}_{\alpha} e_{\alpha i}+\frac{1}{2}(F_i^{I}+F_{ext,i})\delta_t,
\end{eqnarray}
or, more obviously, directly from the components of the moment space vector, $\widehat{\mbox{\boldmath$\bar{f}$}}$,
i.e., $\widehat{\bar{f}}_1$ and $\widehat{\bar{f}}_{\beta}$, where $\beta=4,6,8$ for density and momentum,
respectively.

\section{Macroscopic Dynamical Equations for Multiphase Flow}
\label{mrt3macrod}
In this section, the macroscopic dynamical equations for the 3D MRT LBE multiphase flow model
developed above will be derived by employing the Chapman-Enskog multiscale analysis~\cite{chapman64} for
the D3Q15 velocity model. The macroscopic equations
for other 3D MRT models, such as the D3Q19 model discussed in the previous section, can be found in an
analogous way.
Introducing the expansions~\cite{he97}
\begin{eqnarray}
f_{\alpha}( \overrightarrow{x}+\overrightarrow{e_{\alpha}}\delta_t,t+\delta_t )&=&
\sum_{n=0}^{\infty} \frac{\epsilon^n}{n!} D_{t_n}^nf_{\alpha}(\overrightarrow{x},t),\\
 D_{t_n} &\equiv& \partial_{t_n}+e_{\alpha k} \partial_k,\\
 f_{\alpha}&=&\sum_{n=0}^{\infty}\epsilon^n f_{\alpha}^{(n)},\\
\partial_t&=&\sum_{n=0}^{\infty}\epsilon^n \partial_{t_n},
\end{eqnarray}
where $\epsilon=\delta_t$ in Eq. (\ref{eq:mrt3lbe}), the following equations are obtained as consecutive
orders of the parameter $\epsilon$:
\begin{eqnarray}
O(\epsilon^0): f_{\alpha}^{(0)}&=&f_{\alpha}^{eq}\label{eq:order0mrt3},\\
O(\epsilon^1): D_{t_0} f_{\alpha}^{(0)}&=&-\sum_{\beta}\Lambda_{\alpha \beta} f_{\beta}^{(1)}+S_{\alpha}\label{eq:order1mrt3},\\
O(\epsilon^2): \partial_{t_1} f_{\alpha}^{(0)}+ D_{t_0}\left(I_{\alpha\beta}-\frac{1}{2}\Lambda_{\alpha\beta}\right)
                        f_{\beta}^{(1)}&=&-\sum_{\beta}\Lambda_{\alpha\beta} f_{\beta}^{(2)}.
\label{eq:order2mrt3}
\end{eqnarray}
In equivalent moment space, obtained by multiplying Eqs.(\ref{eq:order0mrt3})-(\ref{eq:order2mrt3})
by the transformation matrix
$\mathcal{T}$, they become
\begin{eqnarray}
O(\epsilon^0): \mbox{\boldmath$\widehat{f}$}^{(0)}&=&\mbox{\boldmath$\widehat{f}$}^{eq}\label{eq:morder0mrt3},\\
O(\epsilon^1): \left(\partial_{t_0}+ \widehat{\mathcal{E}}_i \partial_i \right) \mbox{\boldmath$\widehat{f}$}^{(0)}&=&
                -  \widehat{\Lambda}  \mbox{\boldmath$\widehat{f}$}^{(1)} +
        \mbox{\boldmath$\widehat{\varsigma}$}\label{eq:morder1mrt3},\\
O(\epsilon^2): \partial_{t_1} \mbox{\boldmath$\widehat{f}$}^{(0)}+
                              \left(\partial_{t_0}+   \widehat{\mathcal{E}}_i \partial_i\right)
                  \left( \mathcal{I}   -\frac{1}{2} \widehat{\Lambda} \right)
                  \mbox{\boldmath$\widehat{f}$}^{(1)}&=&
                 -  \widehat{\Lambda} \mbox{\boldmath$\widehat{f}$}^{(2)},
\label{eq:morder2mrt3}
\end{eqnarray}
where $ \widehat{\mathcal{E}}_i =\mathcal{T} e_{\alpha i}  \mathcal{T}^{-1}$.

First let us simplify the source term $S_{\alpha}$ in Eq.(\ref{eq:order1mrt3}) as
\begin{eqnarray}
S_{\alpha}      &=& w_{\alpha}\left[ \frac{3}{c^2}\left(e_{\alpha x}-u_x\right)+
            \frac{9}{c^4}\left(  \overrightarrow{e_{\alpha}} \cdot  \overrightarrow{u} \right)e_{\alpha x}\right]F_x+   \nonumber\\
        & & w_{\alpha}\left[ \frac{3}{c^2}\left(e_{\alpha y}-u_y\right)+
            \frac{9}{c^4}\left(  \overrightarrow{e_{\alpha}} \cdot  \overrightarrow{u} \right)e_{\alpha y}\right]F_y+   \nonumber\\
        & & w_{\alpha}\left[ \frac{3}{c^2}\left(e_{\alpha z}-u_z\right)+
            \frac{9}{c^4}\left(  \overrightarrow{e_{\alpha}} \cdot  \overrightarrow{u} \right)e_{\alpha z}\right]F_z
\label{eq:simpleforce}
\end{eqnarray}
by neglecting terms of the order of $O(Ma^2)$ or higher in its definition, i.e. Eq.(\ref{eq:source3mrt}) and
Eq. (\ref{eq:trunceqmrt3}). In Eq. (\ref{eq:simpleforce}), $F_x$, $F_y$ and $F_z$ are the Cartesian components
of the net force experienced by the particle populations including those that lead to phase segregation and
surface tension effects and imposed forces (e.g., $F_x=-\partial_x\psi+F_{s,x}+F_{ext,x}$).

Upon substituting Eq. (\ref{eq:simpleforce}), after pre-multiplying by $\mathcal{T}$,
in Eq. (\ref{eq:morder1mrt3}), we get the components of the
\emph{first-order equations} in moment space, i.e.
\begin{equation}
\partial_{t_0} \rho + \partial_x j_x + \partial_y j_y + \partial_z j_z=0,
\label{eq:mfirstorder1}
\end{equation}
\begin{eqnarray}
& &
\partial_{t_0} \left(-\rho+\frac{1}{\rho}  \overrightarrow{j} \cdot \overrightarrow{j}  \right)
-\frac{1}{3}\left(\partial_x j_x + \partial_y j_y + \partial_z j_z\right)= \nonumber \\
& &
-s_2 e^{(1)}+2\left(F_xu_x+F_yu_y+F_zu_z\right),
\label{eq:mfirstorder2}
\end{eqnarray}
\begin{eqnarray}
& &
\partial_{t_0} \left(\rho-\frac{5}{\rho}  \overrightarrow{j} \cdot  \overrightarrow{j} \right)
-\frac{7}{3}\left(\partial_x j_x + \partial_y j_y + \partial_z j_z\right)= \nonumber \\
& &
-s_3 e^{2(1)}-10\left(F_xu_x+F_yu_y+F_zu_z\right),
\label{eq:mfirstorder3}
\end{eqnarray}
\begin{equation}
\partial_{t_0} j_x + \partial_x \left(\frac{1}{3}\rho+\frac{j_x^2}{\rho}\right) +
             \partial_y \left(                \frac{j_xj_y}{\rho}\right)+
             \partial_z \left(                \frac{j_xj_z}{\rho}\right)= F_x,
\label{eq:mfirstorder4}
\end{equation}
\begin{eqnarray}
& &
\partial_{t_0} \left(-\frac{7}{3}j_x\right) +
\partial_x\left(-\frac{7}{9}\rho + \frac{1}{3\rho}\left[-7j_x^2+5j_y^2+5j_z^2  \right] \right)+ \nonumber \\
& &
\partial_y\left(\frac{j_xj_y}{\rho}\right) + \partial_z\left(\frac{j_xj_z}{\rho}\right)=-s_5 q_x^{(1)}-\frac{7}{3}F_x,
\label{eq:mfirstorder5}
\end{eqnarray}
\begin{equation}
\partial_{t_0} j_y + \partial_x \left(                \frac{j_xj_y}{\rho}\right) +
             \partial_y \left(\frac{1}{3}\rho+\frac{j_y^2}{\rho}\right)+
             \partial_z \left(                \frac{j_yj_z}{\rho}\right)= F_y,
\label{eq:mfirstorder6}
\end{equation}
\begin{eqnarray}
& &
\partial_{t_0} \left(-\frac{7}{3}j_y\right) +
\partial_x\left(\frac{j_xj_y}{\rho}\right)
\partial_y\left(-\frac{7}{9}\rho + \frac{1}{3\rho}\left[5j_x^2-7j_y^2+5j_z^2  \right] \right)+ \nonumber \\
& &
\partial_z\left(\frac{j_yj_z}{\rho}\right)=-s_7 q_y^{(1)}-\frac{7}{3}F_y,
\label{eq:mfirstorder7}
\end{eqnarray}
\begin{equation}
\partial_{t_0} j_z + \partial_x \left(                \frac{j_xj_z}{\rho}\right) +
             \partial_y \left(                \frac{j_yj_z}{\rho}\right)+
             \partial_z \left(\frac{1}{3}\rho+\frac{j_z^2}{\rho}\right)= F_z,
\label{eq:mfirstorder8}
\end{equation}
\begin{eqnarray}
& &
\partial_{t_0} \left(-\frac{7}{3}j_z\right) +
\partial_x\left(\frac{j_xj_z}{\rho}\right)+
\partial_y\left(\frac{j_yj_z}{\rho}\right)+ \nonumber \\
& &
\partial_z\left(-\frac{7}{9}\rho + \frac{1}{3\rho}\left[5j_x^2+5j_y^2-7j_z^2  \right] \right)=
-s_9 q_z^{(1)}-\frac{7}{3}F_z,
\label{eq:mfirstorder9}
\end{eqnarray}
\begin{eqnarray}
& &
\partial_{t_0} \left(\frac{1}{\rho}\left[2j_x^2-\left(j_y^2+j_z^2\right)\right]\right) +
\frac{2}{3}\left[\partial_x(2j_x)-\partial_yj_y-\partial_zj_z\right]=\nonumber\\
& &
-3s_{10} p_{xx}^{(1)}+2\left(2F_xu_x-F_yu_y-F_zu_z\right),
\label{eq:mfirstorder10}
\end{eqnarray}
\begin{eqnarray}
& &
\partial_{t_0} \left(\frac{1}{\rho}\left[j_y^2-j_z^2\right]\right) +
\frac{2}{3}\left[\partial_yj_y-\partial_zj_z\right]=\nonumber\\
& &
-3s_{11} p_{ww}^{(1)}+2\left(2F_yu_y-F_zu_z\right),
\label{eq:mfirstorder11}
\end{eqnarray}
\begin{equation}
\partial_{t_0} \left(\frac{1}{\rho}j_xj_y\right) +
\frac{1}{3}\left[\partial_xj_y+\partial_yj_x\right]=
-3s_{12} p_{xy}^{(1)}+F_xu_y+F_yu_x,
\label{eq:mfirstorder12}
\end{equation}
\begin{equation}
\partial_{t_0} \left(\frac{1}{\rho}j_yj_z\right) +
\frac{1}{3}\left[\partial_yj_z+\partial_zj_y\right]=
-3s_{13} p_{yz}^{(1)}+F_yu_z+F_zu_y,
\label{eq:mfirstorder13}
\end{equation}
\begin{equation}
\partial_{t_0} \left(\frac{1}{\rho}j_xj_z\right) +
\frac{1}{3}\left[\partial_xj_z+\partial_zj_x\right]=
-3s_{14} p_{zx}^{(1)}+F_xu_z+F_zu_x,
\label{eq:mfirstorder14}
\end{equation}
and
\begin{equation}
\partial_x\left(\frac{j_yj_z}{\rho}\right)+
\partial_y\left(\frac{j_xj_z}{\rho}\right)+
\partial_z\left(\frac{j_xj_y}{\rho}\right)=-s_{15}m_{xyz}^{(1)}.
\label{eq:mfirstorder15}
\end{equation}
The evolution equations express variations at the $t_{0}$ time scale level,
the dynamical relationships between various moments constructed from the velocity space,
their relaxation parameters and the net forces that include those that lead to phase
segregation and surface tension effects and any external forces acting on the particle populations.

Similarly, the components of the \emph{second-order equations} in moment space, i.e.
Eq. (\ref{eq:morder2mrt3}), can be obtained. Our interest is in the dynamical
equations for the conserved moments. For brevity, here we express below only the second order equations
of the conserved moments representing variations at the $t_{1}$ time scale.
\begin{equation}
\partial_{t_1} \rho =0,
\label{eq:msecondorder1}
\end{equation}
\begin{eqnarray}
& &
\partial_{t_1} j_x+
\partial_x \left(\frac{1}{3}\left[1-\frac{1}{2}s_2\right]e^{(1)}+
                \left[1-\frac{1}{2}s_{10}\right]p_{xx}^{(1)}\right)+\nonumber\\
& &
\partial_y \left(\left[1-\frac{1}{2}s_{12}\right]p_{xy}^{(1)}\right)+
\partial_z \left(\left[1-\frac{1}{2}s_{14}\right]p_{zx}^{(1)}\right)=0,
\label{eq:msecondorder4}
\end{eqnarray}
\begin{eqnarray}
& &
\partial_{t_1} j_y+
\partial_x \left(\left[1-\frac{1}{2}s_{12}\right]p_{xy}^{(1)}\right)+ \nonumber\\
& &
\partial_y \left(\frac{1}{3}\left[1-\frac{1}{2}s_{2}\right]e^{(1)}-
             \frac{1}{2}\left[1-\frac{1}{2}s_{10}\right]p_{xx}^{(1)}+
             \frac{1}{2}\left[1-\frac{1}{2}s_{11}\right]p_{ww}^{(1)}
           \right)+ \nonumber\\
& &
\partial_z \left(\left[1-\frac{1}{2}s_{13}\right]p_{yz}^{(1)}\right)=0,
\label{eq:msecondorder6}
\end{eqnarray}
and
\begin{eqnarray}
& &
\partial_{t_1} j_z+
\partial_x \left(\left[1-\frac{1}{2}s_{14}\right]p_{zx}^{(1)}\right)+
\partial_y \left(\left[1-\frac{1}{2}s_{13}\right]p_{yz}^{(1)}\right)+ \nonumber\\
& &
\partial_z \left(\frac{1}{3}\left[1-\frac{1}{2}s_{2}\right]e^{(1)}-
             \frac{1}{2}\left[1-\frac{1}{2}s_{10}\right]p_{xx}^{(1)}-
             \frac{1}{2}\left[1-\frac{1}{2}s_{11}\right]p_{ww}^{(1)}
           \right)=0.
\label{eq:msecondorder8}
\end{eqnarray}

Now, combining the first- and second-order equations for the conserved moments,
i.e. Eqs. (\ref{eq:mfirstorder1}) and (\ref{eq:msecondorder1}),
     Eqs. (\ref{eq:mfirstorder4}) and (\ref{eq:msecondorder4}),
     Eqs. (\ref{eq:mfirstorder6}) and (\ref{eq:msecondorder6}) and
     Eqs. (\ref{eq:mfirstorder8}) and (\ref{eq:msecondorder8}), by using
     $\partial_t=\partial_{t_0}+\epsilon\partial_{t_1}$, we get
\begin{equation}
\partial_t \rho + \partial_x j_x + \partial_y j_y + \partial_z j_z=0,
\label{eq:conteqnmrt3}
\end{equation}
\begin{eqnarray}
& &
\partial_t j_x+
\partial_x \left(\frac{1}{3}\rho+\frac{1}{\rho}j_x^2+
         \frac{1}{3}\epsilon\left[1-\frac{1}{2}s_2\right]e^{(1)} \right. + \nonumber\\
& &
       \left. \epsilon\left[1-\frac{1}{2}s_{10}\right]p_{xx}^{(1)}\right)+
\partial_y \left( \frac{j_xj_y}{\rho}+ \epsilon\left[1-\frac{1}{2}s_{12}\right]p_{xy}^{(1)}\right)+ \nonumber\\
& &
\partial_z \left( \frac{j_xj_z}{\rho}+ \epsilon\left[1-\frac{1}{2}s_{14}\right]p_{zx}^{(1)}\right)=F_x,
\label{eq:momxeqnmrt3}
\end{eqnarray}
\begin{eqnarray}
& &
\partial_t j_y+
\partial_x \left(\frac{j_xj_y}{\rho}+
                    \epsilon\left[1-\frac{1}{2}s_{12}\right]p_{xy}^{(1)} \right) + \nonumber\\
& &
\partial_y \left(\frac{1}{3}\rho+ \frac{1}{\rho}j_y^2+
         \epsilon\frac{1}{3}\left[1-\frac{1}{2}s_2\right]e^{(1)}-
         \epsilon\frac{1}{2}\left[1-\frac{1}{2}s_{10}\right]p_{xx}^{(1)} \right. \nonumber\\
& &
       \left. \epsilon\frac{1}{2}\left[1-\frac{1}{2}s_{11}\right]p_{ww}^{(1)} \right)+
\partial_z \left( \frac{j_xj_z}{\rho}+ \epsilon\left[1-\frac{1}{2}s_{13}\right]p_{yz}^{(1)}\right)=F_y,
\label{eq:momyeqnmrt3}
\end{eqnarray}
and
\begin{eqnarray}
& &
\partial_t j_z+
\partial_x \left(\frac{j_xj_z}{\rho}+
                    \epsilon\left[1-\frac{1}{2}s_{14}\right]p_{zx}^{(1)} \right) + \nonumber\\
& &
\partial_y \left(\frac{j_yj_z}{\rho}+
         \epsilon\left[1-\frac{1}{2}s_{13}\right]p_{yz}^{(1)}\right)+
\partial_z \left(\frac{1}{3}\rho+\frac{1}{\rho}j_z^2+  \right. \nonumber\\
& &
       \left. \frac{1}{3}\epsilon\left[1-\frac{1}{2}s_2\right]e^{(1)}-
              \frac{1}{2}\epsilon\left[1-\frac{1}{2}s_{10}\right]p_{xx}^{(1)}-
              \frac{1}{2}\epsilon\left[1-\frac{1}{2}s_{11}\right]p_{ww}^{(1)}
       \right)=F_z,
\label{eq:momzeqnmrt3}
\end{eqnarray}
respectively. In these equations, $e^{(1)}$ and $p_{xx}^{(1)}$, $p_{ww}^{(1)}$, $p_{xy}^{(1)}$,
$p_{yz}^{(1)}$ and $p_{zx}^{(1)}$ are unknowns to be determined. Writing expressions for
these terms from Eqs. (\ref{eq:mfirstorder2}) and  Eqs. (\ref{eq:mfirstorder10})-(\ref{eq:mfirstorder14}),
respectively, and employing the continuity and momentum equations, i.e., Eqs. (\ref{eq:mfirstorder1}),
 (\ref{eq:mfirstorder4}), (\ref{eq:mfirstorder6}), (\ref{eq:mfirstorder8}), and neglecting terms of
order $O(Ma^3)$ or higher, we get
\begin{eqnarray}
e^{(1)}&=&-\frac{2}{3}\frac{1}{s_2}\left(\partial_xj_x+\partial_yj_y+\partial_zj_z\right)=
        -\frac{2}{3}\frac{1}{s_2} \overrightarrow{\nabla} \cdot \overrightarrow{j} \approx e-e^{eq} ,
\label{eq:e1eqn}\\
p_{xx}^{(1)}&=&-\frac{2}{9}\frac{1}{s_{10}}\left(2\partial_xj_x-\partial_yj_y-\partial_zj_z\right) \approx p_{xx}-p_{xx}^{eq} ,
\label{eq:pxx1eqn}\\
p_{ww}^{(1)}&=&-\frac{2}{3}\frac{1}{s_{11}}\left(\partial_yj_y-\partial_zj_z\right) \approx p_{ww}-p_{ww}^{eq} ,
\label{eq:pww1eqn}\\
p_{xy}^{(1)}&=&-\frac{1}{3}\frac{1}{s_{12}}\left(\partial_xj_y+\partial_yj_x\right) \approx p_{xy}-p_{xy}^{eq} ,
\label{eq:pxy1eqn}\\
p_{yz}^{(1)}&=&-\frac{1}{3}\frac{1}{s_{13}}\left(\partial_yj_z+\partial_zj_y\right) \approx p_{yz}-p_{yz}^{eq} ,
\label{eq:pyz1eqn}\\
p_{zx}^{(1)}&=&-\frac{1}{3}\frac{1}{s_{14}}\left(\partial_xj_z+\partial_zj_x\right) \approx p_{zx}-p_{zx}^{eq} .
\label{eq:pzx1eqn}
\end{eqnarray}
These equations represent the various gradients of the mass flux fields or momentum in terms of
non-equilibrium parts of certain moments and their corresponding relaxation times.

Using Eqs. (\ref{eq:e1eqn})-(\ref{eq:pzx1eqn}) in Eqs. (\ref{eq:momxeqnmrt3})-(\ref{eq:momzeqnmrt3}), the
momentum equations simplify to
\begin{eqnarray}
& &
\partial_t j_x+\partial_x\left(\frac{j_x^2}{\rho}\right)+
           \partial_y\left(\frac{j_yj_x}{\rho}\right)+
           \partial_z\left(\frac{j_zj_x}{\rho}\right)=\nonumber\\
& &
-\partial_x\left(\frac{1}{3}\rho\right)+F_x+
\partial_x\left(2\nu\left[\partial_xj_x-\frac{1}{3}  \overrightarrow{\nabla} \cdot   \overrightarrow{j} \right]+
        \zeta  \overrightarrow{\nabla} \cdot   \overrightarrow{j} \right)+ \nonumber\\
& &
\partial_y\left(\nu\left(\partial_xj_y+\partial_yj_x\right)\right)+
\partial_z\left(\nu\left(\partial_xj_z+\partial_zj_x\right)\right),
\label{eq:momx1eqnmrt3}
\end{eqnarray}
\begin{eqnarray}
& &
\partial_t j_y+\partial_x\left(\frac{j_xj_y}{\rho}\right)+
           \partial_y\left(\frac{j_y^2}{\rho}\right)+
           \partial_z\left(\frac{j_zj_y}{\rho}\right)=\nonumber\\
& &
-\partial_y\left(\frac{1}{3}\rho\right)+F_y+
\partial_x\left(\nu\left(\partial_xj_y+\partial_yj_x\right)\right)+\nonumber\\
& &
\partial_y\left(2\nu\left[\partial_yj_y-\frac{1}{3}  \overrightarrow{\nabla} \cdot   \overrightarrow{j} \right]+
        \zeta  \overrightarrow{\nabla} \cdot  \overrightarrow{j} \right)+
\partial_z\left(\nu\left(\partial_yj_z+\partial_zj_y\right)\right),
\label{eq:momy1eqnmrt3}
\end{eqnarray}
and
\begin{eqnarray}
& &
\partial_t j_z+\partial_x\left(\frac{j_xj_z}{\rho}\right)+
           \partial_y\left(\frac{j_yj_z}{\rho}\right)+
           \partial_z\left(\frac{j_z^2}{\rho}\right)=\nonumber\\
& &
-\partial_z\left(\frac{1}{3}\rho\right)+F_z+
\partial_x\left(\nu\left(\partial_xj_z+\partial_zj_x\right)\right)+\nonumber\\
& &
\partial_y\left(\nu\left(\partial_yj_z+\partial_zj_y\right)\right)+
\partial_z\left(2\nu\left[\partial_zj_z-\frac{1}{3}  \overrightarrow{\nabla} \cdot   \overrightarrow{j} \right]+
        \zeta  \overrightarrow{\nabla} \cdot  \overrightarrow{j} \right).
\label{eq:momz1eqnmrt3}
\end{eqnarray}
where $\zeta$ and $\nu$ are the kinematic bulk and shear viscosities repectively. These are
related to the relaxation parameters of the energy and the stress tensor moments by
\begin{eqnarray}
\zeta&=&\frac{2}{9}\left(\frac{1}{s_{2}}-\frac{1}{2}\right)\delta_t,
\label{eq:bulkvmrt3}\\
\nu&=&\frac{1}{3}\left(\frac{1}{s_{\beta}}-\frac{1}{2}\right)\delta_t, \beta=10,11,12,13,14.
\label{eq:shearvmrt3}
\end{eqnarray}
Notice that some of the relaxation parameters ($s_{10}=s_{11}=\cdots=s_{14}$) should be equal to
one another to maintain the isotropy of the viscous stress tensor. From
Eqs. (\ref{eq:bulkvmrt3}) and (\ref{eq:shearvmrt3}), we see that the bulk and shear viscosities
can be chosen independently. In particular,
this allows us to choose a bulk viscosity larger than the shear viscosity, so that acoustic modes are
attenuated quickly. This is helpful in certain physical situations and could also aid in improving numerical
stability. Finally, substituting for the forcing terms, i.e. $F_i=-\partial_i \psi + F_{s,i}+F_{ext,i}$,
and recognizing that the pressure is given by $P=\psi+1/3\rho$, the macroscopic dynamical equations for the
conserved moments, i.e. density and momentum, are given by
\begin{eqnarray}
\partial_t \rho+  \overrightarrow{\nabla} \cdot  \left(\rho   \overrightarrow{u} \right)&=&0,\label{multi1}\\
\rho\left[\partial_t u_x+  \overrightarrow{u} \cdot  \overrightarrow{\nabla} u_x\right]&=&
\partial_x P+F_{s,x}+F_{ext,x}+\partial_x \sigma_{xx}^{v}+\partial_y \sigma_{xy}^{v}+\partial_z \sigma_{xz}^{v}, \label{multi2}\\
\rho\left[\partial_t u_y+  \overrightarrow{u} \cdot  \overrightarrow{\nabla} u_y\right]&=&
\partial_y P+F_{s,y}+F_{ext,y}+\partial_x \sigma_{yx}^{v}+\partial_y \sigma_{yy}^{v}+\partial_z \sigma_{yz}^{v}, \label{multi3}\\
\rho\left[\partial_t u_z+  \overrightarrow{u} \cdot  \overrightarrow{\nabla} u_z\right]&=&
\partial_z P+F_{s,z}+F_{ext,z}+\partial_x \sigma_{zx}^{v}+\partial_y \sigma_{zy}^{v}+\partial_z \sigma_{zz}^{v}\label{multi4},
\end{eqnarray}
where $\sigma_{ij}^{v}$ is the deviatoric part of the viscous stress tensor,
\begin{equation}
\sigma_{ij}^{v}=\nu\left[\left(\partial_j\left(\rho u_i\right)+\partial_i\left(\rho u_j\right)\right)
                     -\frac{2}{3}  \overrightarrow{\nabla} \cdot  \left(\rho  \overrightarrow{u} \right)  \delta_{ij} \right]+
\zeta   \overrightarrow{\nabla} \cdot  \left(\rho  \overrightarrow{u} \right)  \delta_{ij},
\end{equation}
$F_{s,i}$ is the surface tension force given in Eq. (\ref{eq:forcempmrt3}), $F_{ext,i}$ is the external force,
and $i,j\in\left\{x,y,z\right\}$. The equations (\ref{multi1})-(\ref{multi4}), indeed, correspond to the
macroscopic equations for multiphase flow derived from kinetic theory~\cite{zou99}.

\section{3D MRT LBE Model for Multiphase Flow with Reduced Compressibility Effects}
\label{mrt3modelc}
Multiphase LBE models based on inter-particle interactions~\cite{he98a}
face difficulties for fluids far from the critical point and/or in the presence of external forces~\cite{he04}. This
difficulty is related to the calculation of the inter-particle force involving the term
$\partial_j \psi$, in the model which becomes quite large across interfaces. As a remedy, He and co-workers~\cite{he99}
developed a suitable transformation of the distribution function  $f_{\alpha}$, by invoking the
incompressibility condition of the fluid, to $g_{\alpha}$ that determines the hydrodynamic fields, i.e. pressure
and velocity fields. A separate distribution function is also constructed to capture the interface through an
order parameter, which is different from density. The introduction of these features helps in reducing compressibility
effects in multiphase problems. In this section, we apply this idea to the 3D MRT model discussed in Section \ref{mrt3model}.
Following He \emph{et al.}~\cite{he99}, we replace the distribution function $f_{\alpha}$ by another distribution
function $g_{\alpha}$ through the transformation
\begin{equation}
g_{\alpha}=f_{\alpha}RT+\psi(\rho)\frac{f_{\alpha}^{eq,M}(\rho,\overrightarrow{0} )}{\rho}.
\label{eq:transgmrt3}
\end{equation}
By considering the fluid to be incompressible, i.e.
\begin{equation}
\frac{d}{dt}\psi(\rho)=\left( \partial_t+u_k \partial_k \right)\psi (\rho)=0,
\end{equation}
and using Eqs. (\ref{eq:transgmrt3}) and (\ref{eq:implicittr1mrt3}),
Eq. (\ref{eq:mrt3lbe}) becomes
\begin{eqnarray}
& &\bar{g}_{\alpha}( \overrightarrow{x} +  \overrightarrow{e_{\alpha}} \delta_t,t+\delta_t )-
\bar{g}_{\alpha}(   \overrightarrow{x} ,t )= \nonumber\\
& &
-\sum_{\beta} \Lambda_{\alpha \beta}\left( \bar{g}_{\beta}-g_{\beta}^{eq} \right)|_{(x,t)}+
 \sum_{\beta} \left(I_{\alpha\beta}-\frac{1}{2}\Lambda_{\alpha\beta}\right)S_{g\beta}|_{(x,t)}\delta_t,
\label{eq:lbee2mrt3}
\end{eqnarray}
where
\begin{equation}
g_{\alpha}^{eq}=f_{\alpha}^{eq}RT+\psi(\rho)\frac{f_{\alpha}^{eq,M}(\rho,\overrightarrow{0} )}{\rho}.
\end{equation}
The corresponding source terms are
\begin{eqnarray}
S_{g\alpha}&=&(e_{\alpha j}-u_j)\times
               \left[
              (F_j^{I}+F_{ext,j})\frac{f_{\alpha}^{eq,M}(\rho,  \overrightarrow{u} )}{\rho}- \right. \nonumber \\
           & & \left. \left(
          \frac{f_{\alpha}^{eq,M}(\rho,  \overrightarrow{u} )}{\rho}-\frac{f_{\alpha}^{eq,M}(\rho,\overrightarrow{0} )}{\rho}
          \right)\partial_j \psi(\rho)
          \right].
\label{eq:sourcegmrt3}
\end{eqnarray}
Equations (\ref{eq:lbee2mrt3})-(\ref{eq:sourcegmrt3}) are numerically more stable than
the original formulation since the term $\partial_j\psi$ is multiplied by a factor
proportional to the Mach number $O(Ma)$, instead of being $O(1)$ when the "incompressible" transformation
is not employed. As a result, this term becomes smaller in the incompressible limit. Now, applying the
transformation matrix to the above system, we get
\begin{eqnarray}
& &\widehat{\bar{g}}_{\alpha}( \overrightarrow{x} +  \overrightarrow{e_{\alpha}} \delta_t,t+\delta_t )-
\widehat{\bar{g}}_{\alpha}(   \overrightarrow{x} ,t )= \nonumber \\
& &
-\sum_{\beta} \widehat{\Lambda}_{\alpha \beta}\left( \widehat{\bar{g}}_{\beta}-\widehat{g}_{\beta}^{eq} \right)|_{(x,t)}+
 \sum_{\beta} \left(I_{\alpha\beta}-\frac{1}{2}\widehat{\Lambda}_{\alpha\beta}\right)\widehat{S}_{g\beta}|_{(x,t)}\delta_t.
 \label{eq:lbee3mrt3}
 \end{eqnarray}

In this new framework, we still need to introduce an order parameter to capture interfaces.
Here, we employ a function $\phi$ referred to henceforth as the index function, in place of the density as
the order parameter. The evolution equation of the distribution function, whose emergent dynamics govern the
index function, must have the term responsible for maintaining phase segregation and mass conservation.
In this regard, we use Eqs. (\ref{eq:mrt3lbe}) and (\ref{eq:source3mrt}), by keeping the term involving $\partial_j\psi$
and dropping the rest as they play no role in mass conservation. In addition, the density is replaced by the
index function in these equations. Hence, the evolution of the distribution function for the index function
is given by
\begin{eqnarray}
\bar{f}_{\alpha}(   \overrightarrow{x} +  \overrightarrow{e_{\alpha}} \delta_t,t+\delta_t )-
\bar{f}_{\alpha}(   \overrightarrow{x} ,t )&=&
-\sum_{\beta} \Lambda_{\alpha \beta}\left( \bar{f}_{\beta}-\frac{\phi}{\rho}f_{\beta}^{eq} \right)|_{(x,t)}+ \nonumber\\
& &
 \sum_{\beta} \left(I_{\alpha\beta}-\frac{1}{2}\Lambda_{\alpha\beta}\right)S_{f\beta}|_{(x,t)}\delta_t,
\label{eq:lbee2mrt13}
\end{eqnarray}
where
\begin{equation}
S_{f\alpha}=\frac{(e_j-u_j)(-\partial_j \psi(\phi))}{\rho RT}f_{\alpha}^{eq,M}(\rho,  \overrightarrow{u} ).
\end{equation}
In moment space, this becomes
\begin{eqnarray}
& &\widehat{\bar{f}}_{\alpha}(   \overrightarrow{x} +  \overrightarrow{e_{\alpha}} \delta_t,t+\delta_t )-
\widehat{\bar{f}}_{\alpha}(   \overrightarrow{x} ,t )= \nonumber \\
& &
-\sum_{\beta} \widehat{\Lambda}_{\alpha \beta}\left( \widehat{\bar{f}}_{\beta}-\frac{\phi}{\rho}\widehat{f}_{\beta}^{eq} \right)|_{(x,t)}+
 \sum_{\beta} \left(I_{\alpha\beta}-\frac{1}{2}\widehat{\Lambda}_{\alpha\beta}\right)\widehat{S}_{f\beta}|_{(x,t)}\delta_t.
\label{eq:lbee3mrt13}
\end{eqnarray}

It follows that hydrodynamical variables, such as the pressure and fluid velocity, can be
obtained by taking appropriate kinetic moments of the distribution function $g_{\alpha}$, i.e.
\begin{eqnarray}
P&=&\sum_{\alpha} \bar{g}_{\alpha}-\frac{1}{2}u_j\partial_j \psi(\rho),\\
\rho RT u_i &=& \sum_{\alpha} \bar{g}_{\alpha} e_{\alpha i}+\frac{1}{2}RT\left( F_{s,i}+F_{ext,i} \right)\delta_t.
\end{eqnarray}
The index function is obtained from the distribution function $\bar{f}_{\alpha}$ by taking the zeroth
kinetic moment, i.e.
\begin{equation}
\phi=\sum_{\alpha}\bar{f}_{\alpha}.
\end{equation}
The density is obtained from the index function through linear interpolation, i.e.
\begin{equation}
\rho(\phi)= \rho_L+\frac{\phi-\phi_L}{\phi_H-\phi_L}(\rho_H-\rho_L),
\label{eq:rhointp}
\end{equation}
where $\rho_L$ and $\rho_H$ are densities of light and heavy fluids, respectively, and $\phi_L$ and
$\phi_H$ refer to the minimum and maximum values of the index function, respectively. These limits of the index
function are determined from Maxwell's equal area construction~\cite{rowlinson82} applied to the function
$\psi(\phi)+\phi RT$. If the viscosities are different in different phases, the corresponding relaxation parameters
($s_{11}$ through $s_{14}$, which are equal to one another) in each phase are obtained from Eq. (\ref{eq:shearvmrt3})
and their variation across interfaces are determined through a linear interpolation similar to that for the
density as shown in Eq. (\ref{eq:rhointp}).

The numerical implementation of this MRT model is as follows: In the collision step, the evolution of distribution
functions are computed in \emph{moment space} ($\mbox{\boldmath$\widehat{f}$}$ and $\mbox{\boldmath$\widehat{g}$}$)
through direct relaxation of moments at rates determined by the diagonal collision matrix $\widehat{\Lambda}$ with
appropriate updates from the source terms. After transforming the post-collision moments back to the velocity space,
the streaming step is performed in \emph{velocity space} ($\mbox{\boldmath$f$}$ and $\mbox{\boldmath$g$}$).
After these two computational steps, the various macroscopic fields are updated. This allows an efficient
implementation of the MRT model for multiphase problems.

In terms of the computational memory and time resources, the MRT model requires a negligible increase in memory and
a moderate increase in computational time, when compared to the BGK model, if appropriate optimization strategies are
taken. Depending on the lattice velocity model, the transformation matrix $\mathcal{T}$ is either a
$15 \times 15$ or $19 \times 19$ matrix. These matrices are the same throughout the domain and so need not be stored
at every lattice site, thus contributing to negligible increase in memory. The corresponding relaxation time collision
matrices $\widehat{\Lambda}$ in moment space which are diagonal matrices with $15$ or $19$ elements, respectively,
also do not contribute to any significant increase in memory requirement. The equilibrium distributions in moment
space depend only on the local macroscopic field variables and can be computed locally without requiring additional storage.
The memory requirement for the distribution functions for the collision and streaming steps in the MRT model is about
the same as that required for the corresponding BGK version.

As noted in Ref.~\cite{dhumieres02}, suitable code optimization techniques should be applied to avoid substantial
increase in computational time due to the use of the MRT model. First, direct matrix computations, which can lead to
increased computational overhead, should not be carried out in the transformation between velocity
and moment spaces. Instead, the transformation between the components of the distribution functions in velocity
space $f_{\alpha}$ and the moments $\widehat{f}_{\beta}$ should be carried out explicitly using the expressions obtained
from the transformation matrix, which is an integer matrix with several common elements and many zeroes. In these
calculations, all the common sub-expressions should be computed just once. Second, the computational algorithm
should be carried out as discussed in the previous paragraph. With such optimization strategies, the MRT model
only incurs a moderate increase in computational time, about $35\%$ more than the BGK model for the cases discussed
in the next section, but with significantly improved stability.

\section{Results and Discussion}
\label{mrt3benchmarks}
The 3D MRT model developed in the previous section will now be evaluated for some benchmark multiphase problems. Unless otherwise
specified, we express the results in lattice units, i.e., the velocities are scaled by the particle velocity $c$
and the distance by the lattice spacing $\delta_x$. We first consider a static multiphase problem, namely, the verification of
the well-known Laplace-Young relation for a stationary 3D drop. This relation expresses the force balance between the excess pressure
inside the drop and the surface tension force. If $\Delta P$ is the pressure difference between the inside and outside of a drop of
radius $R_d$ with surface tension $\sigma$, we have $\Delta P=2 \sigma/R_d$.

We first simulate this problem with a D3Q15 lattice by considering drops of three different radii $15$, $21$ and $27$.
The corresponding 3D domain is discretized by $41\times 41 \times 41$, $61\times 61 \times 61$, and $81\times 81 \times 81$
lattice sites, respectively. Periodic boundary conditions are considered in all directions. We simulate drops in a gas with an equal shear
kinematic viscosity of $1.0\times 10^{-2}$ for the gas and liquid and a density ratio of $4$. We have chosen equal kinematic viscosities
for both the phases just for simplicity. The MRT model developed in this work is not limited to choosing only identical viscosities for
both the phases. Indeed, as noted in the previous section,  when there is a viscosity mismatch, the corresponding relaxation parameters will be
different in each phase and their values across interfaces can be obtained by linear interpolation of viscosities through the index
function $\phi$. It is a known limitation that the underlying LB interfacial physical model~\cite{he98a,he99} is stable for only low density
ratios. The MRT model has relatively little influence on this limitation. However, as will be shown below, it can significantly increase the
stability at lower viscosities.

It may be noted that when the BGK model is employed to simulate multiphase flow with the viscosity used above,
the computations become unstable. As discussed in Section \ref{mrt3model}, we choose the relaxation parameters
$s_1$, $s_4$, $s_6$ and $s_8$ to be $1.0$. The parameters $s_{\alpha}$, where $\alpha=10, 11, 12, 13, 14$, as shown
in Eq. (\ref{eq:shearvmrt3}), are obtained from the shear kinematic viscosity. On the other hand, the bulk kinematic viscosity is related to
$s_2$ (see Eq.(\ref{eq:bulkvmrt3})). We consider $s_2=1.0$. Thus, the time scales for the shear and bulk viscosities are different.
The remaining free parameters $s_3$, $s_5$, $s_7$, $s_9$ and $s_{15}$, whose values have no effect on hydrodynamics, are chosen
to be equal to $1.0$ for simplicity.

\begin{figure}
\begin{center}
\includegraphics[height=3.25in,clip=]{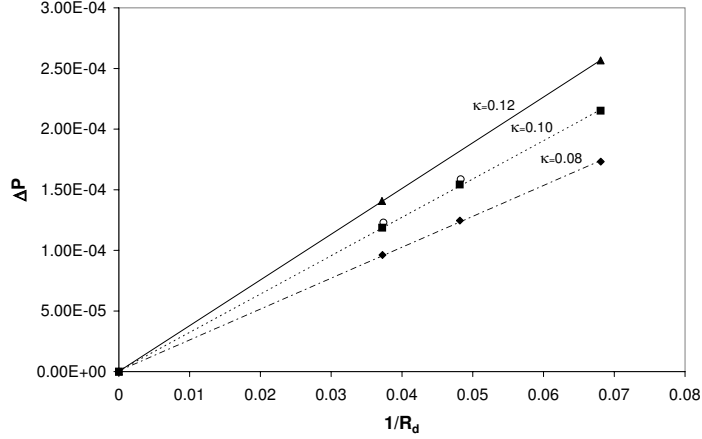}
\caption{Pressure difference across a drop as a function of radius for different values of the surface tension
     parameter ($\kappa$); Bold symbols employ the D3Q15 model, open symbol the D3Q19 model.}
\label{fig:laplaceyoung}
\end{center}
\end{figure}

Figure~\ref{fig:laplaceyoung} shows the computed pressure difference across the drop interface as a function of inverse
drop radii for three values of the surface tension parameter: $\kappa=0.08$, $0.10$ and $0.12$. The lines correspond to a
linear fit for the computed data. We find that for these selected parameters the computed pressure difference agrees with the
Laplace-Young relation prediction to within $8\%$. To check the validity of the MRT lattice model with a larger velocity set,
i.e. the D3Q19 model, we performed some selected computations, which are shown as open symbols in the same figure.
The D3Q15 and D3Q19 models yield results that have negligible difference between them. Henceforth, for comparison purposes
we will only consider results obtained with the D3Q15 lattice. The density profile across the drop interface
is found to be characterized well with no anisotropic effects. As with other LB models, and with methods based on the direct
solution of Navier-stokes equations for multiphase flows, velocity currents around the interfaces are observed.
These are generally small and is found to be proportional to the surface tension parameter $\kappa$.

Next, we consider a dynamical problem, namely, the oscillation of a liquid drop immersed in a gas. We
employ the analytical solution of Miller and Scriven (1968)~\cite{miller68}
for comparison with the computed time periods.
According to Ref.~\cite{miller68}, the frequency of the $n^{th}$ mode of oscillation of a drop is given by
\begin{equation}
\omega_{n}=\omega_{n}^{*}-\frac{1}{2} \chi \omega_{n}^{*\frac{1}{2}}+\frac{1}{4}\chi^2,
\label{eq:msperiod}
\end{equation}
where $\omega_{n}$ is the angular response frequency, and $\omega_{n}^{*}$ is Lamb's natural resonance
frequency expressed as~\cite{lamb32}
\begin{equation}
\left(\omega_{n}^{*}\right)^{2}=
\frac{n(n+1)(n-1)(n+2)}{R_d^3 \left[ n\rho_{g}+(n+1)\rho_{l} \right]} \sigma.
\end{equation}
$R_d$ is the equilibrium radius of the drop, $\sigma$ is the interfacial surface tension, and $\rho_{l}$ and
$\rho_{g}$ are the densities of the two fluids. The parameter $\chi$ is given by
\begin{equation}
\chi =
\frac{(2n+1)^2 (\mu_{l} \mu_{g} \rho_{l} \rho_{g})^{\frac{1}{2}}}
{2^{\frac{1}{2}} R_d \left[ n\rho_{g}+(n+1)\rho_{l} \right]
\left[ (\mu_{l} \rho_{l})^{\frac{1}{2}}+ (\mu_{g} \rho_{g})^{\frac{1}{2}} \right]},
\end{equation}
where $\mu_{l}$ and $\mu_{g}$ are the dynamic viscosities of the two fluids. Here, the subscripts $g$ and $l$
refer to the gas and liquid phases, respectively. We consider the second mode of oscillation, i.e. $n=2$, and
analytical expression for the time period $T_{anal}$ is obtained from Eq. (\ref{eq:msperiod}) through
$T_{anal}=2\pi/\omega_{2}$.

\begin{figure}
\begin{center}
\includegraphics[height=5.00in,clip=]{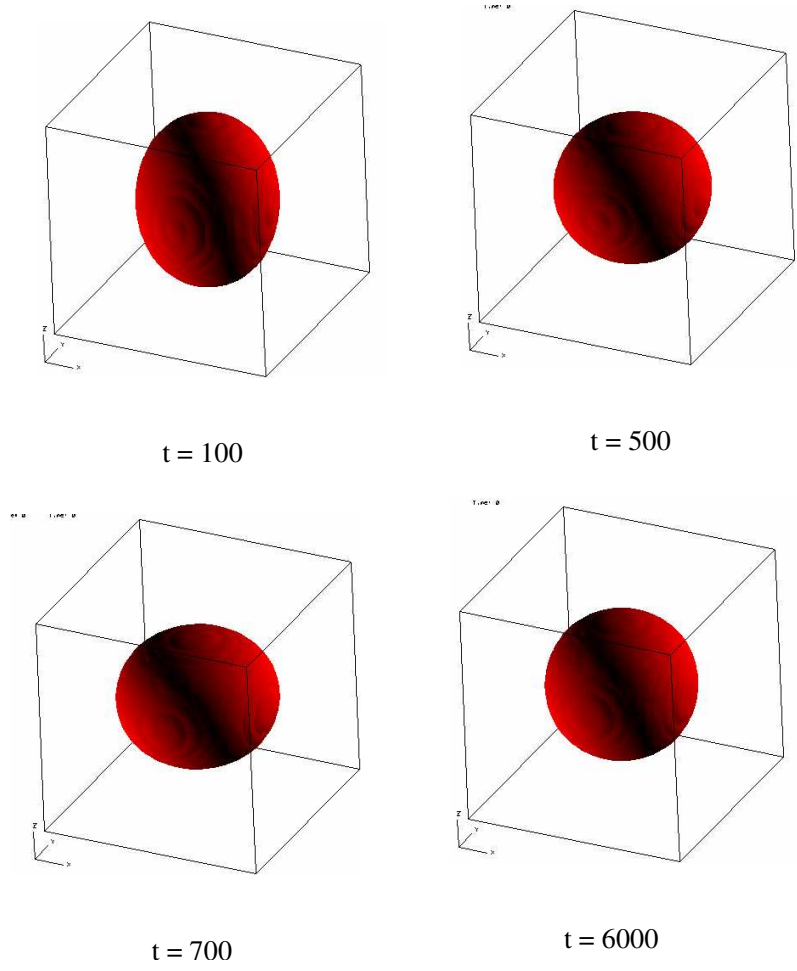}
\caption{Configurations of an oscillating drop at different times in lattice units;
     $\rho_l/\rho_g=4$,$\mu_l/\mu_g=4$,$\nu_l=\nu_g=1.6667\times 10^{-2}$, $\kappa=0.1$.}
\label{fig:drop_config}
\end{center}
\end{figure}

The initial computational setup consists of a prolate spheroid with a minimum and maximum radii
of $11$ and $15$, respectively, and placed in the center of a domain discretized by $41\times 41 \times 41$ lattices. We
consider a density ratio of $4$ for different values of shear viscosity and surface tension parameters. Figure~\ref{fig:drop_config}
shows the drop configurations at different times for $\nu=1.6667\times 10^{-2}$ and $\kappa=0.10$.
At $t=100$, the
drop has a prolate shape which temporarily assumes an approximate spherical shape at $t=500$. At $t=700$, notice that the drop
becomes an oblate spheroid and at the longer time of $t=6000$, the drop assumes its equilibrium spherical configuration. Computations
are also performed by reducing the shear viscosity by $2.5$ and $5$ as a function of time
from the above value, i.e. to
$6.667\times 10^{-3}$ and $3.333\times 10^{-3}$. When the BGK model is employed for computations of
drop oscillations, it is stable only to the point where the viscosity is lowered to  $\nu=1.6667\times 10^{-2}$.
Thus, the 3D MRT multiphase flow model allows the viscosity to be lowered (or enhances the Reynolds number) by a factor of $5$ in this case
as compared to the BGK model employing the same underlying physical model, based on the kinetic theory of dense fluids~\cite{he98a,he99}.

\begin{figure}
\begin{center}
\includegraphics[height=3.50in,clip=]{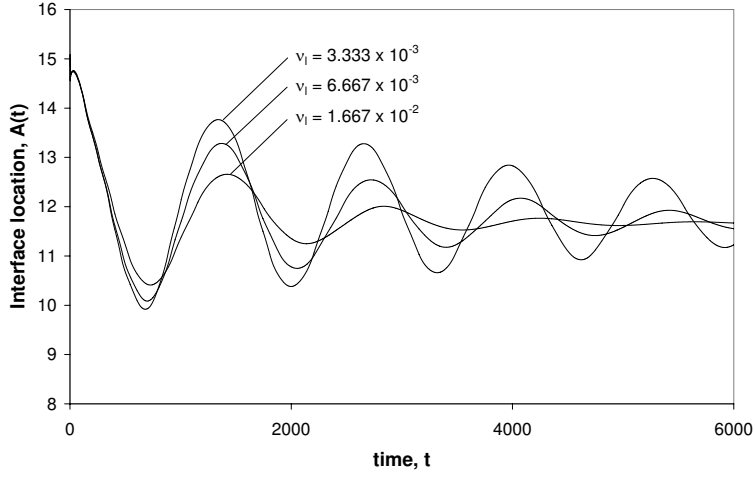}
\caption{Interface location of an oscillating drop as a function of time for different kinematic viscosities, $\nu_l$;
     $\rho_l/\rho_g=4$,$\mu_l/\mu_g=4$, $\kappa=0.1$.}
\label{fig:dropint1}
\end{center}
\end{figure}

Figure~\ref{fig:dropint1} shows the interface locations of the oscillating drops as a function of time
for the three shear viscosities noted above when $\kappa=0.1$.
As expected, when the viscosity is reduced, it takes longer for the drops to reach the equilibrium shape by viscous dissipation.
The computed time periods of oscillations $T_{LBE}$ are $1201$, $1173$ and $1159$ when the viscosities are
$\nu=1.6667\times 10^{-2}$, $6.667\times 10^{-3}$ and $3.333\times 10^{-3}$, respectively. The corresponding analytical time periods
$T_{anal}$ are $1258$, $1202$ and $1173$, respectively. Thus the maximum relative error is $4.80\%$. Let us now decrease
the surface tension parameter $\kappa$ to $0.08$ from $0.10$. Decreasing the surface tension parameter reduces the surface
tension of the drop. As a result, it is expected that the drop would take longer to complete a period of oscillation.

\begin{figure}
\begin{center}
\includegraphics[height=3.50in,clip=]{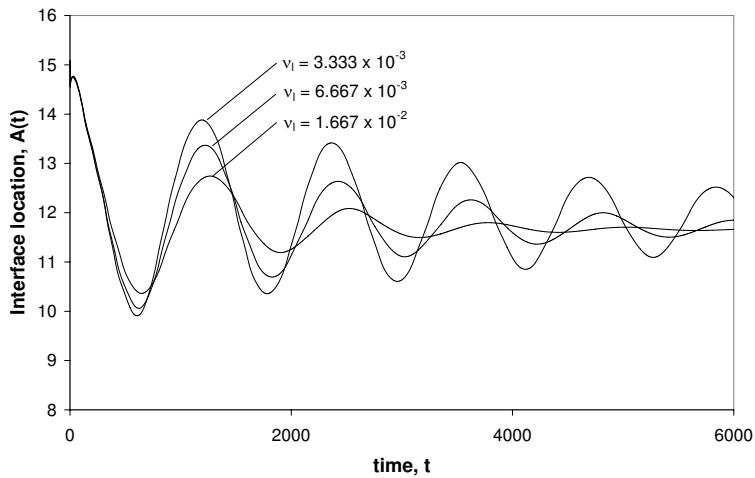}
\caption{Interface location of an oscillating drop as a function of time for different kinematic viscosities, $\nu_l$;
     $\rho_l/\rho_g=4$,$\mu_l/\mu_g=4$, $\kappa=0.08$.}
\label{fig:dropint2}
\end{center}
\end{figure}

\begin{figure}
\begin{center}
\includegraphics[height=3.50in,clip=]{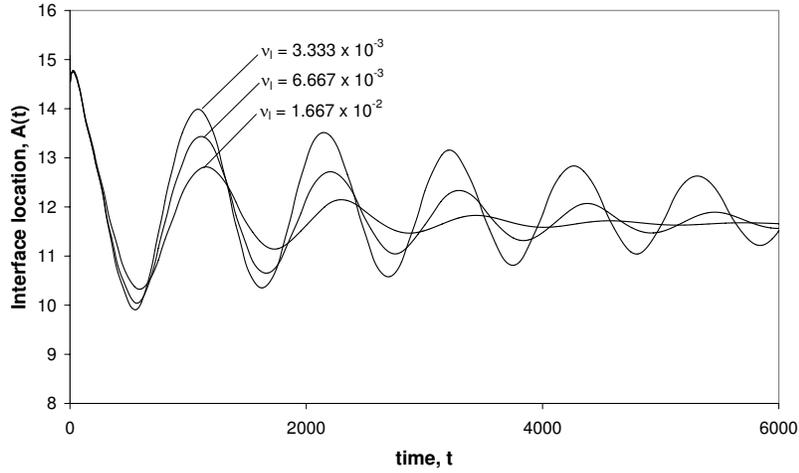}
\caption{Interface location of an oscillating drop as a function of time for different kinematic viscosities, $\nu_l$;
     $\rho_l/\rho_g=4$,$\mu_l/\mu_g=4$, $\kappa=0.12$.}
\label{fig:dropint3}
\end{center}
\end{figure}

Figure~\ref{fig:dropint2}
shows the interface locations for the three shear viscosities noted above when $\kappa=0.08$. The computed
time period, for example, when $\nu=1.6667\times 10^{-2}$ is $1412$ and the corresponding analytical value is $1348$. Both these values
are higher than those at the same value of viscosity for the higher surface tension parameter, which is consistent with expectation.
To confirm the trend with surface tension, we performed computations by increasing $\kappa$ to $0.12$ from $0.10$.
Figure~\ref{fig:dropint3}
shows the interface locations when $\kappa=0.12$. When $\nu=1.6667\times 10^{-2}$, the computed time period
is $1143$, which agrees with the analytical value within $4.60\%$. As this value is lower than with  the surface
tension parameter of $0.1$, we conclude that the computations reproduce variations with surface tension which are
consistent with expectation from the analytical solution. The 3D MRT model is able to reproduce the time period of
oscillations within $5\%$.

The 3D MRT model remains as stable as in the previous cases even for
more complex dynamical problems such as when drops undergo shearing
and other types of motions. In a recent study, this model was
applied to simulate collision of a pair of drops under different
conditions~\cite{premnath05a}. In particular, when employed to study
off-center collisions it is able to reproduce experimentally
observed~\cite{qian97} complex interfacial shape changes due to such
dynamical events as shearing, rotation, stretching deformation and
breakup, while maintaining an improved stability compared to the BGK
model. Indeed, in some cases reported in Ref.~\cite{premnath05a} as
much as an order of magnitude improvement in stability was observed
by the use of MRT model in lieu of the BGK model.

\section{Summary and Conclusions}
\label{summary}
In this paper, we develop 3D MRT LB models for multiphase flows. They employ the LBE with a generalized
collision term together with forcing terms representing interfacial physics.
By considering 3D lattice velocity models and a second-order discretization of the forcing terms, the LBE
written in terms of the distribution function is transformed into an equivalent system represented by a set
of conserved and non-conserved moments. The conserved or hydrodynamic moments include the fluid density
and momentum, while the non-conserved or kinetic moments include heat fluxes and viscous stresses.
The models are developed such that when the collision term in the LBE is written in moment
space, the moments relax to their equilibrium values at rates that can be adjusted independently.

By applying the Chapman-Enskog multiscale analysis to the MRT
models, we show that they correctly recover the 3D hydrodynamical
equations for multiphase flows in the continuum limit. The dynamical
relationships between various moments and the forcing terms,
including surface tension forces, are systematically derived. Some
of the relaxation parameters in the collision term are shown to be
related to the shear and bulk kinematic viscosities of the fluid.
The ability to independently adjust the relaxation parameters
enhances the stability of the MRT models. The models are evaluated
for accuracy by solving some multiphase test problems. It is shown
that the 3D MRT models verify the Laplace-Young relation for static
drops to within $8\%$. Computations with MRT models of drop
oscillations show that shear viscosities can be lowered by a factor
of $5$ when compared to the BGK model. The computed time period of
oscillations agrees with the analytical solution to within $5\%$.
The MRT model, when applied to more complex multiphase problems such
as binary drop collisions, is found to be as stable as that for
these simple canonical problems. The MRT approach developed in this
work can be extended to LBE multiphase flow models such as those
developed in Refs.~\cite{lee05,zheng06} to handle high density ratio
problems. Moreover, it also provides a more natural framework to
extend the LBE for more general situations such as viscoelastic or
thermal effects in multiphase flows in future investigations.

\newpage

\section*{Acknowledgements}
\label{acknow}
The authors thank Drs. X. He, L.-S. Luo and M.E. McCracken for helpful discussions and the Purdue University Computing
Center (PUCC) and the National Center for Supercomputing Applications (NCSA) for providing computing resources.

\newpage

\appendix
\section{Appendix. Simplified Transformation Matrices for Unit Lattice Spacing}
\label{appendix}
For unit lattice spacing and time steps, i.e. $c=1$, the
transformation matrix, Eq. (\ref{eq:trmatrixd3q15}), for the D3Q15
model simplifies to
\[\mathcal{T}= \left[ \scriptsize
\begin{array}{rrrrrrrrrrrrrrr}
1 & 1 & 1&1 & 1 & 1&1 & 1 & 1&1 & 1 &1&1 & 1 & 1\\[-3mm]
-2 & -1 & -1&-1 & -1 & -1&-1 & 1 & 1&1 & 1 & 1&1 & 1 & 1\\[-3mm]
16 & -4 & -4&-4 & -4 & -4&-4 & 1 & 1&1 &1 & 1&1 & 1 & 1\\[-3mm]
0 & 1 & -1&0 & 0 & 0&0 & 1 & -1&1 &-1 & 1&-1 & 1 & -1\\[-3mm]
0 & -4 & 4&0 & 0 & 0&0 & 1 & -1&1 &-1 & 1&-1 & 1 & -1\\[-3mm]
0 & 0 & 0&1 & -1 & 0&0 & 1 & 1&-1 &-1 & 1&1 & -1 & -1\\[-3mm]
0 & 0 & 0&-4 & 4 & 0&0 & 1 & 1&-1 &-1 & 1&1 & -1 & -1\\[-3mm]
0 & 0 & 0&0 & 0 & 1&-1 & 1 & 1&1 &1 & -1&-1 & -1 & -1\\[-3mm]
0 & 0 & 0&0 & 0 & -4&4 & 1 & 1&1 &1 & -1&-1 & -1 & -1\\[-3mm]
0 & 2 & 2&-1 & -1 & -1&-1 & 0 & 0&0 & 0 &0&0 & 0 & 0\\[-3mm]
0 & 0 & 0&1 & 1 & -1& -1 & 0 & 0&0 & 0 &0&0 & 0 & 0\\[-3mm]
0 & 0 & 0&0 & 0 & 0&0 & 1 & -1&-1 &1 &1&-1 &-1 & 1\\[-3mm]
0 & 0 & 0&0 & 0 & 0&0 & 1 & 1&-1 & -1 &-1&-1 &1 & 1\\[-3mm]
0 & 0 & 0&0 & 0 & 0&0 & 1 & -1&1 & -1 &-1&1 & -1 & 1\\[-3mm]
0 & 0 & 0&0 & 0 & 0&0 & 1 & -1&-1 & 1 &-1&1 & 1 & -1
\end{array}\right] \] \vspace*{50mm}%

\pagebreak

and that for the D3Q19 model to
\[\mathcal{T}= \left[ \scriptsize
\begin{array}{rrrrrrrrrrrrrrrrrrr}
1 & 1 & 1&1 & 1 & 1&1 & 1 & 1&1 & 1 &1&1 & 1 & 1&1 & 1 & 1&1\\[-3mm]
-30 & -11 & -11&-11 & -11 & -11&-11 & 8 & 8&8 & 8 & 8&8 & 8 & 8&8 & 8 & 8&8\\[-3mm]
12 & -4 & -4&-4 & -4 & -4&-4 & 1 & 1&1 &1 & 1&1 & 1 & 1&1 & 1 & 1&1\\[-3mm]
0 & 1 & -1&0 & 0 & 0&0 & 1 & -1&1 &-1 & 1&-1 & 1 & -1&0 & 0&0&0\\[-3mm]
0 & -4 & 4&0 & 0 & 0&0 & 1 & -1&1 &-1 & 1&-1 & 1 & -1&0 & 0&0&0\\[-3mm]
0 & 0 & 0&1 & -1 & 0&0 & 1 & 1&-1 &-1 & 0&0 & 0 & 0&1 & -1&1&-1\\[-3mm]
0 & 0 & 0&-4 & 4 & 0&0 & 1 & 1&-1 &-1 & 0&0 & 0 & 0&1 & -1&1&-1\\[-3mm]
0 & 0 & 0&0 & 0 & 1&-1 & 0 & 0&0 &0 & 1&1 & -1 & -1&1 & 1&-1&-1\\[-3mm]
0 & 0 & 0&0 & 0 & -4&4 & 0 & 0&0 &0 & 1&1 & -1 & -1&1 & 1&-1&-1\\[-3mm]
0 & 2 & 2&-1 & -1 & -1&-1 & 1 & 1&1 & 1 &1&1 & 1 & 1&-2 & -2 & -2& -2\\[-3mm]
0 & -4 & -4&2 & 2 & 2& 2 & 1 & 1&1 & 1 &1&1 & 1 & 1&-2 & -2 & -2& -2\\[-3mm]
0 & 0 & 0&1 & 1 & -1&-1 & 1 & 1&1 & 1 &-1&-1 &-1 & -1&0 & 0 & 0&0\\[-3mm]
0 & 0 & 0&-2 & -2 & 2& 2 & 1 & 1&1 & 1 &-1&-1 &-1 & -1&0 & 0 & 0&0\\[-3mm]
0 & 0 & 0&0 & 0 & 0&0 & 1 & -1&-1 & 1 &0&0 & 0 & 0&0 & 0 & 0&0\\[-3mm]
0 & 0 & 0&0 & 0 & 0&0 & 0 & 0&0 & 0 &0&0 & 0 & 0&1 & -1 & -1&1\\[-3mm]
0 & 0 & 0&0 & 0 & 0&0 & 0& 0&0 & 0 &1&-1 & -1 & 1&0 & 0 & 0&0\\[-3mm]
0 & 0 & 0&0 & 0 & 0&0 & 1& -1&1 & -1 &-1&1 & -1 & 1&0 & 0 & 0&0\\[-3mm]
0 & 0 & 0&0 & 0 & 0&0 & -1& -1&1 & 1 &0&0 & 0 & 0&1 & -1 & 1&-1\\[-3mm]
0 & 0 & 0&0 & 0 & 0&0 & 0& 0&0& 0 &1&1 & -1 & -1&-1 & -1 & 1&1
\end{array}\right] \]

\end{document}